\documentclass[amsmath,amssymb,amsfonts,aps,pre,twocolumn,longbibliography]{revtex4-1}

\usepackage[english]{babel}

\usepackage[utf8]{inputenc}
\usepackage[T1]{fontenc}
\usepackage{lmodern}

\usepackage{epsfig}
\usepackage{color}
\usepackage[dvipsnames]{xcolor}
\usepackage{hyperref} 
\usepackage[caption=false,labelformat=simple]{subfig}
\usepackage{mathbbol}

\makeatletter
\renewcommand{\fnum@figure}{FIG. \thefigure}
\makeatother

\begin{document}
\renewcommand{\appendixname}{APPENDIX}

\title{Pair approximation for the $q$-voter models with quenched disorder on networks}

\author{Arkadiusz J\k{e}drzejewski}
\email{arkadiusz.jedrzejewski@pwr.edu.pl}
\affiliation{Department of Theoretical Physics, Wroc\l{}aw University of Science and Technology, Wroc\l{}aw, Poland}
\author{Katarzyna Sznajd-Weron} 
\affiliation{Department of Theoretical Physics, Wroc\l{}aw University of Science and Technology, Wroc\l{}aw, Poland}

\date{\today}

\begin{abstract}
Using two models of opinion dynamics, the $q$-voter model with independence and the $q$-voter model with anticonformity, we discuss how the change of disorder from annealed to quenched affects phase transitions on networks.
To derive phase diagrams on networks, we develop the pair approximation for the quenched versions of the models. 
This formalism can be also applied to other quenched dynamics of similar kind.
The results indicate that such a change of disorder eliminates all discontinuous phase transitions and broadens ordered phases.
We show that although the annealed and quenched types of disorder lead to the same result in the $q$-voter model with anticonformity at the mean-field level, they do lead to distinct phase diagrams on networks.
These phase diagrams shift towards each other as the average node degree of a network increases, and eventually, they coincide in the mean-field limit. 
In contrast, for the $q$-voter model with independence, the phase diagrams move towards the same direction regardless of the disorder type, and they do not coincide even in the mean-field limit.
To validate our results, we carry out Monte Carlo simulations on random regular graphs and Barab\'{a}si-Albert networks. 
Although the pair approximation may incorrectly predict the type of phase transitions for the annealed models, we have not observed such errors for their quenched counterparts.\\\\
Post-print of \href{https://doi.org/10.1103/PhysRevE.105.064306}{Phys. Rev. E \textbf{105}, 064306 (2022)}.\\
Copyright (2022) by the American Physical Society.
\end{abstract}

\maketitle
\section{introduction}
The question of how the type of disorder impacts phase transitions is relevant to various dynamical systems that rely on random processes \cite{Dor:Gol:Men:08, Cas:For:Lor:09,Mun:etal:10,Man:etal:21,Odo:Sim:21}.
In agent-based models of opinion dynamics, annealed and quenched disorders can be associated with different approaches to modeling human behavior \cite{Szn:Szw:Wer:14,Che:Gal:18,Jed:Szn:19}.
Annealed disorder reflects the situation in which the behavior of agents is probabilistic and can change in time during the evolution of the system, whereas quenched disorder corresponds to the behavior of agents that is fixed in time although may vary between agents.
In low-dimensional systems, the presence of quenched disorder leads to the elimination of discontinuous phase transitions known as a rounding effect \cite{Aiz:Weh:90, Bor:Mar:Mun:13, Vil:Bon:Mun:14}.
In the mean-field regime, on the other hand, discontinuous phase transitions may survive the introduction of quenched disorder \cite{Jed:Szn:20,Now:Sto:Szn:21}, or they may appear after its introduction \cite{Now:Szn:21}. The change of disorder may also not impact the phase transitions at all \cite{Jed:Szn:17}.
However, neither regular lattices nor mean-field analyses are best suited to describe social systems with their network-like structure \cite{Boc:etal:06}.
Therefore, in this study, we consider networks and discuss how quenched disorder impacts phase transitions displayed by two simple models of opinion dynamics with competing conforming and nonconforming interactions.
The first one is the $q$-voter model with independence, whereas the second one is the $q$-voter model with anticonformity \cite{Nyc:Szn:13,Nyc:Szn:Cis:12}.
Both these models are modifications of the nonlinear $q$-voter model \cite{Cas:Mun:Pas:09} and were introduced to explore how different types of nonconformity impact phase transitions in a well-mixed population \cite{Nyc:Szn:Cis:12}.
Our work extends this study by taking into account the network structure and quenched disorder.

The interplay between conformity and nonconformity is used to study social phenomena in many agent-based models \cite{Gra:Rus:20}.
In particular, agents with fixed behavior classified as either conformity or nonconformity can be found in studies on the $q$-voter model \cite{Jav:Squ:15,Jed:Szn:17,Kha:Tor:19,Now:Sto:Szn:21}, the majority-vote model \cite{Jav:14,Vil:etal:19,Kra:18}, or the Galam model \cite{Sta:Sa:04}.
Quenched disorder may be also related with the interactions between agents  \cite{Bar:21a,Bar:21b,Kra:20,Kra:21}.
However, there are fewer studies in which the annealed and quenched approaches are directly compared within the same dynamics, especially on networks.

How disorder changes phase transitions displayed by the $q$-voter models with nonconformity is well described at the mean-field level \cite{Jed:Szn:17}.
In the $q$-voter model with independence, the change from annealed to quenched disorder eliminates all discontinuous phase transitions and broadens the ordered phases of the continuous ones.
Interestingly, in the $q$-voter model with anticonformity, which displays only continuous phase transitions under the annealed approach, such a change of disorder does not impact the phase transitions at all.
The question which naturally arises is whether the above results remain valid on networks.
Although both the models have been studied extensively under the annealed approach on networks, by the use of the pair approximation and Monte Carlo simulations \cite{Jed:17,Per:etal:18,Gra:Kra:20,Jed:etal:20,Abr:Szn:20,Vie:etal:20,Abr:etal:21}, such a comprehensive analysis is missing for their quenched counterparts. 
If the quenched models are considered on networks, studies rely only on numerical simulations \cite{Szn:Szw:Wer:14,Jav:Squ:15}.

Therefore, in this work, we develop the pair approximation for the quenched versions of the $q$-voter models with different types of nonconformity in order to obtain analytically their phase diagrams on networks. 
The predictions of the pair approximation are validated by conducting Monte Carlo simulations on random regular graphs and Barab\'{a}si-Albert networks.
The obtained results under the quenched approach are compared with the results obtained under the annealed one.
Our comparative analysis on networks reveals differences between the models that are not displayed in well-mixed populations.
This demonstrates how important the network structure is in answering the question about the differences between dynamics.

\section{Annealed and quenched models}
\label{sec:model}
We consider an undirected network with $N$ nodes.
The network represents a social structure where nodes are voters, whereas links indicate relationships between them.
Each node can be in two states: $j\in\{1,-1\}$, or equivalently $j\in\{\uparrow,\downarrow\}$ for the sake of clarity in the further notation. 
These states represent different opinions, suppose a positive and a negative one.
In this setting, one node after another is selected randomly.
This node will have a chance to change its opinion.
Next, we randomly choose $q$ its nearest neighbors without repetition.
They form a group of influence that tries to exert social pressure on the selected voter.
The following steps of the dynamics depend on the specific variant of the $q$-voter model:
\begin{itemize}
    \item \textbf{Annealed approach:} The selected voter behaves as a nonconformist with probability $p$, otherwise, with complementary probability $1-p$, it behaves as a conformist. 
    This approach is considered in Refs.~\cite{Nyc:Szn:Cis:12, Nyc:Szn:13,Szn:Szw:Wer:14,Jed:Szn:17, Now:Sto:Szn:21,Abr:etal:21}. 
    \item \textbf{Quenched approach:} The behavior of the selected voter is predetermined by its personal trait assigned randomly at the beginning of the simulation.
    On average, the fraction of voters that are always nonconformists is $p$, whereas the fraction of voters that are always conformists is $1-p$. 
    This approach is considered in Refs.~\cite{Szn:Szw:Wer:14,Jed:Szn:17,Jav:Squ:15, Now:Sto:Szn:21}.
\end{itemize}
Under the above two approaches, we compare the behavior of the $q$-voter models with different types of nonconformity, i.e., independence and anticonformity.
The dynamical rules that reflect the analyzed social responses are as follows \cite{Nyc:Szn:Cis:12,Nyc:Szn:13}:
\begin{itemize}
    \item \textbf{Conformity:} If all $q$ members of the influence group share the same opinion, the selected voter takes the opinion shared by the group. 
    \item \textbf{Independence:} The selected voter resists social pressure and independently of the influence group changes its opinion to the opposite one with probability $1/2$.
    \item \textbf{Anticonformity:} If all $q$ members of the influence group share the same opinion, the selected voter takes the opposite opinion to the one shared by the group. 
\end{itemize}
If the influence group is not unanimous in the case of conformity or anticonformity, the voter stays with their old opinion.
This choice of behavior in the case of disagreement is consistent with the formulation of the $q$-voter models introduced in Ref.~\cite{Nyc:Szn:Cis:12}, which we refer to.
However, in a more general case, one may assume that when the influence group is not unanimous, the voter still have a chance to change their opinion with some probability $\epsilon$. Such an approach was originally considered in the nonlinear $q$-voter model \cite{Cas:Mun:Pas:09}.
Introducing this generalization to the $q$-voter models with nonconformity may result in richer phase diagrams, and it may be an interesting direction for future studies.

Finally, let us note that we consider a random sequential update, as in most studies. In contrast, a synchronous update is implemented in Ref.~\cite{Jav:Squ:15}.

\section{Pair approximation for quenched models}
\label{sec:pair}
The pair approximation is a general technique used to study various dynamics on static \cite{Jed:17,Per:etal:18,Jed:etal:20,Abr:etal:21,Abr:Szn:20,Gra:Kra:20,Vie:etal:20,Sch:Beh:09,Gle:13,Chm:Gra:Kra:18,Kra:20,Per:Tor:20,Kra:21} as well as coevolutionary networks \cite{Dia:San:Egu:14,Min:San:17,Tor:etal:17,Jed:Tor:etal:20,Rad:San:20}.
This method has already been applied to the $q$-voter models with nonconformity under the annealed approach. 
The calculations for  the $q$-voter model with independence can be found in Refs.~\cite{Jed:17,Per:etal:18,Abr:Szn:20,Gra:Kra:20,Vie:etal:20}, whereas for the $q$-voter model with anticonformity can be found in Refs.~\cite{Jed:etal:20,Abr:etal:21}.
In this section, we develop the pair approximation for the quenched versions of the $q$-voter models. 

As indicated in Sec.~\ref{sec:model}, we distinguish between two states $j\in\{\uparrow,\downarrow\}$, which represent positive and negative opinions of voters, and two types of nodes $\tau\in\{\text{c},\text{n}\}$, which refer to voters that are conformists and nonconformists, respectively.
Note that in the case of the $q$-voter model with independence, nonconformists are equivalent to independent individuals ($\text{n}\equiv \text{i}$), whereas in the case of the $q$-voter model with anticonformity, nonconformists are equivalent to anticonformists ($\text{n}\equiv \text{a}$).
Let $c_\tau$ be the concentration of nodes of type $\tau$ with a positive opinion.
Since the fraction of voters that are nonconformists is $p$, and the rest of them are conformists, the overall concentration of voters with a positive opinion is as follows:
\begin{equation}
\label{eq:partition}
	c=pc_\text{n}+(1-p)c_\text{c}.
\end{equation}

\begin{figure}[!t]
    \centering
    \subfloat{\label{fig:notation:a}}
	\subfloat{\label{fig:notation:b}}
    \epsfig{file=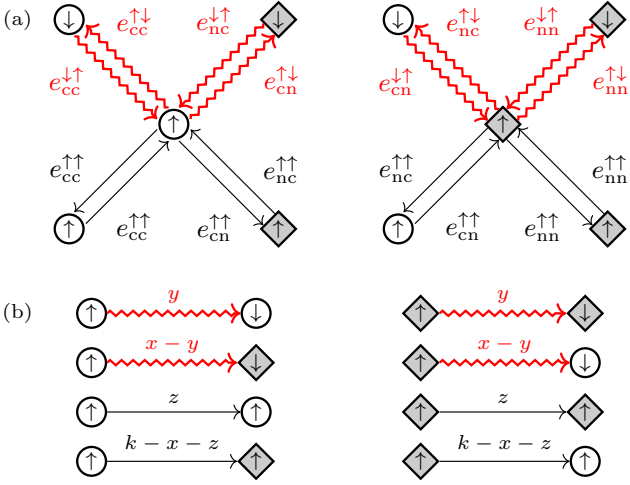,width=\linewidth}
    \caption{\label{fig:notation} Illustration of the used notation. Circles and squares represent different types of nodes, conformists and nonconformists, respectively. An arrow in a node corresponds to its state. Zigzag and straight arrows connecting nodes represent directed links that are active and inactive, respectively. Two cases are presented for which the central node has a positive opinion, and it is either conformist or nonconformist.
    (a) Labels next to the links indicate their classes. (b) Labels above the links indicate the numbers of such out-links belonging to the central node from the graph directly above.}
\end{figure}
To conduct the pair approximation, we use the concept of directed links even though we consider only undirected networks.
This means that we replace each undirected link with two oppositely directed links \cite{Tor:etal:17,Jed:Tor:etal:20}.
Additionally, we group these directed links into classes based on the states and types of nodes they connect.
Let $e_{j_1j_2}^{\tau_1\tau_2}$ denote the concentration of directed links, where the first upper and lower indexes refer to the state and type of the node at the origin of a link, whereas the second indexes refer to the corresponding variables at the end of it.
Note that $e_{j_1j_2}^{\tau_1\tau_2}=e_{j_2j_1}^{\tau_2\tau_1}$ since our networks are undirected.
Figure~\ref{fig:notation:a} illustrates this notation.
Additionally, we have
\begin{align}
	e_{\uparrow\uparrow}^{\text{cc}}&+e_{\downarrow\downarrow}^{\text{cc}}+2e_{\uparrow\downarrow}^{\text{cc}}+e_{\uparrow\uparrow}^{\text{nn}}+e_{\downarrow\downarrow}^{\text{nn}}+2e_{\uparrow\downarrow}^{\text{nn}}\nonumber\\
	&+2e_{\uparrow\uparrow}^{\text{cn}}+2e_{\downarrow\downarrow}^{\text{cn}}+2e_{\uparrow\downarrow}^{\text{cn}}+2e_{\downarrow\uparrow}^{\text{cn}}=1
\end{align}
since this is the sum over all the concentrations of directed links. 
Moreover, since the node types are assigned randomly, the concentrations of directed links connecting nodes of specific types are known:
\begin{align}
e_{\uparrow\uparrow}^{\text{cc}}+e_{\downarrow\downarrow}^{\text{cc}}+2e_{\uparrow\downarrow}^{\text{cc}}=&(1-p)^2,\\
e_{\uparrow\uparrow}^{\text{nn}}+e_{\downarrow\downarrow}^{\text{nn}}+2e_{\uparrow\downarrow}^{\text{nn}}=&p^2,\label{eq:sumeii}\\
e_{\uparrow\uparrow}^{\text{cn}}+e_{\downarrow\downarrow}^{\text{cn}}+e_{\uparrow\downarrow}^{\text{cn}}+e_{\downarrow\uparrow}^{\text{cn}}=&(1-p)p.\label{eq:sumeci}
\end{align}

Let us now consider a randomly selected out-link of a given node.
The first node, at the origin of this link, is of known type and in known state, $\tau_1$ and $j_1$, respectively, whereas the second node, at the end of this link, is of unknown type and in unknown state.
Let $P(j_2|\tau_1,j_1)$ denote the probability that the second node is in state $j_2$. On the other hand, let $P(\tau_2|\tau_1,j_1;j_2)$ denote the conditional probability that the type of the second node is $\tau_2$ given that the state of this node is $j_2$.
The above probabilities are approximated by certain fractions of the concentrations of directed links.
We distinguish between active and inactive links.
Active links refer to the links that connect nodes in different states, whereas inactive links refer to the links that connect nodes in the same states.
First, let us define the probabilities of selecting an active out-link of a node of type $\tau$ that is in state $j$. 
Using our notation, these probabilities are denoted by $P(-j|\tau,j)$, so the explicit formulas are the following (we will also refer to them by Greek letters specified below for brevity of notation):
\begin{align}
    P(\downarrow|\text{c},\uparrow)\equiv&\alpha=\frac{e_{\uparrow\downarrow}^{\text{cc}}+e_{\uparrow\downarrow}^{\text{cn}}}{e_{\uparrow\downarrow}^{\text{cc}}+e_{\uparrow\downarrow}^{\text{cn}}+e_{\uparrow\uparrow}^{\text{cc}}+e_{\uparrow\uparrow}^{\text{cn}}},\label{eq:prob_dcu}\\
    P(\uparrow|\text{c},\downarrow)\equiv&\bar{\alpha}=\frac{e_{\downarrow\uparrow}^{\text{cc}}+e_{\downarrow\uparrow}^{\text{cn}}}{e_{\downarrow\uparrow}^{\text{cc}}+e_{\downarrow\uparrow}^{\text{cn}}+e_{\downarrow\downarrow}^{\text{cc}}+e_{\downarrow\downarrow}^{\text{cn}}},\\
    P(\downarrow|\text{n},\uparrow)\equiv&\widetilde{\alpha}=\frac{e_{\uparrow\downarrow}^{\text{nn}}+e_{\uparrow\downarrow}^{\text{nc}}}{e_{\uparrow\downarrow}^{\text{nn}}+e_{\uparrow\downarrow}^{\text{nc}}+e_{\uparrow\uparrow}^{\text{nc}}+e_{\uparrow\uparrow}^{\text{nn}}},\\
	P(\uparrow|\text{n},\downarrow)\equiv&\widehat{\alpha}=\frac{e_{\downarrow\uparrow}^{\text{nn}}+e_{\downarrow\uparrow}^{\text{nc}}}{e_{\downarrow\uparrow}^{\text{nn}}+e_{\downarrow\uparrow}^{\text{nc}}+e_{\downarrow\downarrow}^{\text{nn}}+e_{\downarrow\downarrow}^{\text{nc}}}.
	\label{eq:prob_uid}
\end{align}
Now, let us define the conditional probabilities that a randomly selected out-link of a node of type $\tau$ that is in state $j$ connects nodes of the same type given that this out-link is active. Using our notation, these probabilities are denoted by $P(\tau|\tau,j;-j)$, so the explicit formulas are the following:
\begin{align}
P(\text{c}|\text{c},\uparrow;\downarrow)\equiv&\beta=\frac{e_{\uparrow\downarrow}^{\text{cc}}}{e_{\uparrow\downarrow}^{\text{cc}}+e_{\uparrow\downarrow}^{\text{cn}}},\label{eq:prob_ccud}\\
P(\text{c}|\text{c},\downarrow;\uparrow)\equiv&\bar{\beta}=\frac{e_{\downarrow\uparrow}^{\text{cc}}}{e_{\downarrow\uparrow}^{\text{cc}}+e_{\downarrow\uparrow}^{\text{cn}}},\\
P(\text{n}|\text{n},\uparrow;\downarrow)\equiv&\widetilde{\beta}=\frac{e_{\uparrow\downarrow}^{\text{nn}}}{e_{\uparrow\downarrow}^{\text{nn}}+e_{\uparrow\downarrow}^{\text{nc}}},\\
P(\text{n}|\text{n},\downarrow;\uparrow)\equiv&\widehat{\beta}=\frac{e_{\downarrow\uparrow}^{\text{nn}}}{e_{\downarrow\uparrow}^{\text{nn}}+e_{\downarrow\uparrow}^{\text{nc}}}.\label{eq:prob_iidu}
\end{align}
Finally, let us define the conditional probabilities that a randomly selected out-link of a node of type $\tau$ that is in state $j$ connects nodes of the same type given that this out-link is inactive. Using our notation, these probabilities are denoted by $P(\tau|\tau,j;j)$, so the explicit formulas are the following:
\begin{align}
P(\text{c}|\text{c},\uparrow;\uparrow)\equiv&\gamma=\frac{e_{\uparrow\uparrow}^{\text{cc}}}{e_{\uparrow\uparrow}^{\text{cc}}+e_{\uparrow\uparrow}^{\text{cn}}},\label{eq:prob_ccuu}\\
P(\text{c}|\text{c},\downarrow;\downarrow)\equiv&\bar{\gamma}=\frac{e_{\downarrow\downarrow}^{\text{cc}}}{e_{\downarrow\downarrow}^{\text{cc}}+e_{\downarrow\downarrow}^{\text{cn}}},\\
P(\text{n}|\text{n},\uparrow;\uparrow)\equiv&\widetilde{\gamma}=\frac{e_{\uparrow\uparrow}^{\text{nn}}}{e_{\uparrow\uparrow}^{\text{nn}}+e_{\uparrow\uparrow}^{\text{nc}}},\\
P(\text{n}|\text{n},\downarrow;\downarrow)\equiv&\widehat{\gamma}=\frac{e_{\downarrow\downarrow}^{\text{nn}}}{e_{\downarrow\downarrow}^{\text{nn}}+e_{\downarrow\downarrow}^{\text{nc}}}.\label{eq:prob_iidd}
\end{align}

Let us consider a node of type $\tau$ that is in state $j$ and has $k$ out-links.
We assume that each of these out-links may be active independently of others with probability $P(-j|\tau,j)$, given by one of Eqs.~(\ref{eq:prob_dcu})-(\ref{eq:prob_uid}).
Thus, the number of active out-links, $x\in\{0,k\}$, is binomially distributed.
Since the node has $x$ active out-links, the remaining $k-x$ out-links are inactive.
Given the number of active out-links $x$, let $f^\tau(x,k)$ denote the probability that a node of type $\tau$ that has $k$ out-links changes its state to the opposite one.
Based on the model definition (see Sec.~\ref{sec:model}), these probabilities for different types of voters are as follows:
\begin{itemize}
    \item conformists:
        \begin{equation}
           f^\text{c}(x,k)=\frac{x!(k-q)!}{k!(x-q)!}\mathbf{1}_{\{x\geq q\}}(x),
        \end{equation}
    \item independent individuals:
        \begin{equation}
           f^\text{i}(x,k)=\frac{1}{2},
        \end{equation}
    \item anticonformists:
        \begin{equation}
           f^\text{a}(x,k)=\frac{(k-x)!(k-q)!}{k!(k-x-q)!}\mathbf{1}_{\{x\leq k-q\}}(x),
        \end{equation}
\end{itemize}
where $\mathbf{1}_A(x)$ is an indicator function defined on a set $A$.
Having defined these quantities, we can write down a differential equation for each concentration $c_\tau$:
\begin{align}
    \frac{dc_\tau}{dt}=&-\sum_{j}c_{j}^\tau\sum_kP^\tau_j(k)\nonumber\\
    &\times\sum^k_x B\big[x;k,P(-j|\tau,j)\big]f^\tau(x,k)j,
    \label{eq:dct}
\end{align} 
where $P^\tau_j(k)$ is the network degree distribution associated only with nodes of type $\tau$ that are in state $j$, the function $B$ returns the binomial probability, i.e.,
\begin{equation}
    B[x;k,\theta]={{k}\choose{x}}\theta^x(1-\theta)^{k-x},
\end{equation}
whereas $c^{\tau}_\uparrow\equiv c_\tau$ and $c^{\tau}_\downarrow\equiv 1-c_\tau$ for notation brevity.
Explicit forms of these equations can be found in Appendix~\ref{sec:apendix}.
Note that Eq.~(\ref{eq:dct}) represents the average change in $c_\tau$ in a time interval $dt=1/N$.
The construction of similar equations for the time evolution of other quantities can be found in Refs.~\cite{Vaz:Egu:08,Jed:17,Jed:Szn:19,Gle:13,Per:etal:18}

Similar differential equations have to be written for each concentration of directed links.
Let us consider further the node from the previous paragraph. 
This node is of type $\tau$, and it is in state $j$.
Moreover, $x$ out of its all $k$ out-links are active.
Now, let us say that among these $x$ active out-links, there are $y\in\{0,x\}$ active out-links that point to nodes of the same type $\tau$.
We assume that each of the active out-links may connect nodes of the same type $\tau$ independently of others with probability $P(\tau|\tau,j;-j)$, given by one of Eqs.~(\ref{eq:prob_ccud})-(\ref{eq:prob_iidu}).
Thus, the number of active out-links that connect nodes of the same type $\tau$, $y$, is binomially distributed.
The remaining $x-y$ active out-links connect nodes of different types.
Analogously, let us say that among $k-x$ inactive out-links, there are $z\in\{0,k-x\}$ inactive out-links that connect nodes of the same type $\tau$.
In the same way, we postulate that this number is binomially distributed with probability $P(\tau|\tau,j;j)$, given by one of Eqs.~(\ref{eq:prob_ccuu})-(\ref{eq:prob_iidd}).
The remaining $k-x-z$ inactive out-links connect nodes of different types. Figure~\ref{fig:notation:b} illustrates how the out-links of a given node are partitioned into different classes according to the above convention.

Having this partition, the differential equations for all the concentrations of directed links have the following form:
\begin{widetext}
\begin{align}
   \frac{d}{dt}e_{j_1j_2}^{\tau_1\tau_2}=&\frac{1}{\langle k\rangle}\sum_{\substack{i\in\{1,2\} \\ j\in\{\uparrow,\downarrow\}}} \left[p\mathbf{1}_{\{\text{n}\}}(\tau_i)+(1-p)\mathbf{1}_{\{\text{c}\}}(\tau_i)\right]c_j^{\tau_i}\sum_k P_j^{\tau_i}(k)\sum_x^k B\big[x;k,P(-j|\tau_i,j)\big]\nonumber\\
   &\times\sum_y^x B\big[y;x,P(\tau_i|\tau_i,j;-j)\big]\sum_z^{k-x}B\big[z;k-x,P(\tau_i|\tau_i,j;j)\big]f^{\tau_i}(x,k)\Delta e_{j_1j_2}^{\tau_1\tau_2}(i,j),
   \label{eq:de}
\end{align}
\end{widetext}
where $\langle k\rangle$ is the average node degree of a network, and $\Delta e_{j_1j_2}^{\tau_1\tau_2}(i,j)$ is given by:
\begin{equation}
    \Delta e_{j_1j_2}^{\tau_1\tau_2}(i,j)=-j_1j_2
    \begin{cases}
    z & \text{if } \tau_1=\tau_{2}\land j=j_{3-i},\\
    -y & \text{if } \tau_1=\tau_{2}\land j\neq j_{3-i},\\
    k-x-z & \text{if } \tau_1\neq\tau_{2}\land j=j_{3-i},\\
    -(x-y) & \text{if } \tau_1\neq\tau_{2}\land j\neq j_{3-i},
    \end{cases}
    \label{eq:deij}
\end{equation}
note that $i\in\{1,2\}$ and $j\in\{\uparrow,\downarrow\}$, as in the sums of the above differential equation.
Equation~(\ref{eq:de}) represents the average change in $e_{j_1j_2}^{\tau_1\tau_2}$ in a time interval $dt=1/N$. The form of Eqs.~(\ref{eq:de}) and (\ref{eq:deij}) follows directly from the introduced link partition convention; see also Fig.~\ref{fig:notation:b}.
In Eq.~(\ref{eq:de}), the formula for the first moment of the binomial distribution can be used to perform the sums after a proper change of the summing indexes \cite{Jed:17}.
The explicit differential equations for all the concentrations of directed links after the summation can be found in Appendix~\ref{sec:apendix}.

In the following section, we present the results for the quenched models based on the above calculations, and we compare them to the results for the annealed models based on the calculations conducted in Refs.~\cite{Jed:17} and \cite{Jed:etal:20}.

\section{Results}
Herein, we present the main results and describe the phase diagrams obtained based on the pair approximation.
The details of Monte Carlo simulations, used to validate our calculations, are presented in Appendix~\ref{sec:apendix-b}.
Throughout the work, we distinguish between two phases---an ordered one and a disordered one.
In the ordered phase, one opinion dominates over the other (i.e., $c\neq0.5$), whereas in the disordered phase, both opinions are equally likely (i.e., $c=0.5$).

\begin{figure}[b!]
	\centering
	\subfloat{\label{fig:phasemap:a}}
	\subfloat{\label{fig:phasemap:b}}
	\subfloat{\label{fig:phasemap:c}}
	\subfloat{\label{fig:phasemap:d}}
	\epsfig{file=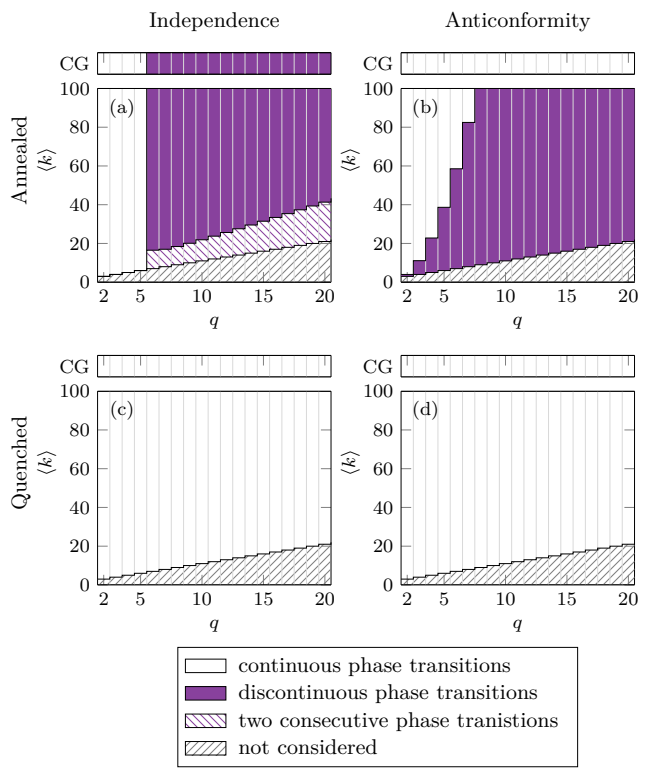,width=\linewidth}
	\caption{\label{fig:phasemap} Types of phase transitions predicted by the pair approximation in $q$-voter model with (a, c) independence (b, d) anticonformity under the (a, b) annealed (c, d) quenched approach. 
	We consider integer values of the average node degree, $\langle k\rangle$, that are bigger than the influence group size, $q$.
	Two consecutive phase transitions refer to continuous and discontinuous phase transitions that happen one after the other, see Figs.~\ref{fig:phasediagram:d} and \ref{fig:phasediagram:e}.
	The top rows labeled by CG correspond to the results on a complete graph of the infinite size, i.e., the case when $\langle k\rangle\to\infty$.}
\end{figure}
\begin{figure*}[!th]
	\centering
	\subfloat{\label{fig:phasediagram:a}}
	\subfloat{\label{fig:phasediagram:b}}
	\subfloat{\label{fig:phasediagram:c}}
	\subfloat{\label{fig:phasediagram:d}}
	\subfloat{\label{fig:phasediagram:e}}
	\subfloat{\label{fig:phasediagram:f}}
	\epsfig{file=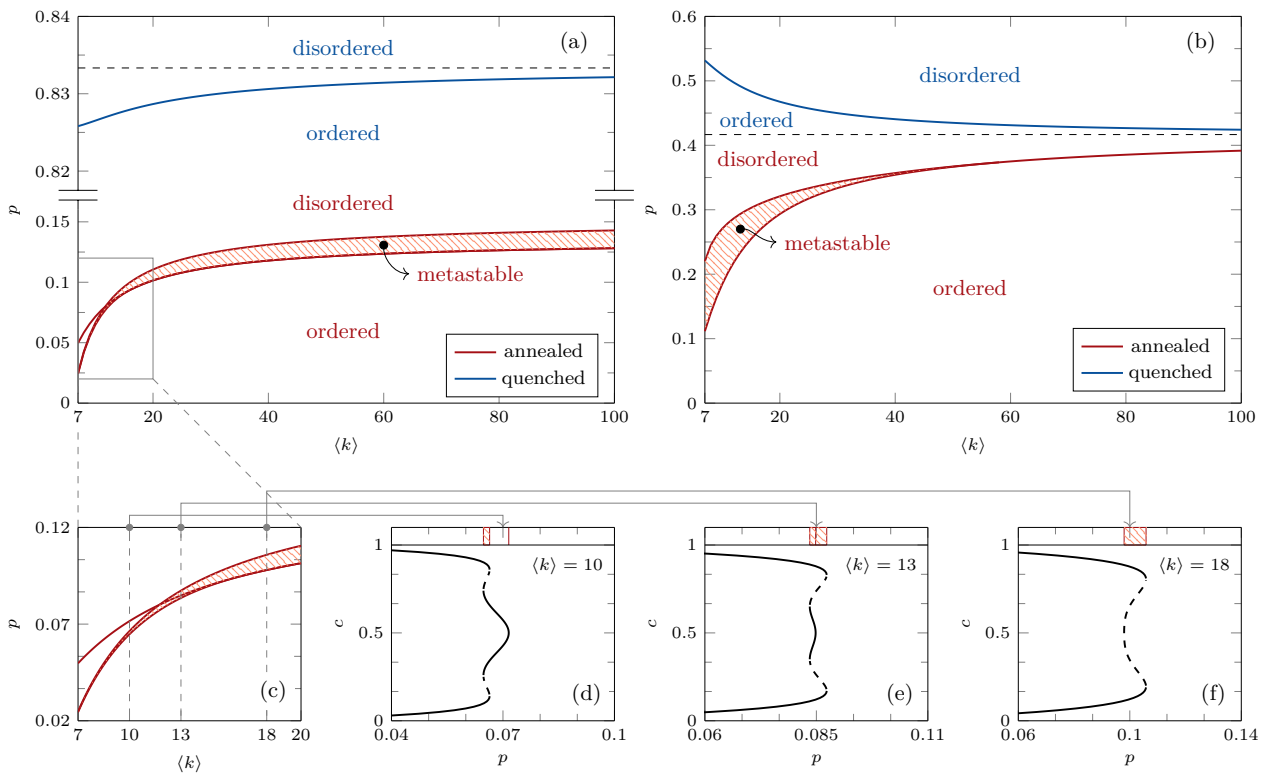,width=\linewidth}
	\caption{\label{fig:phasediagram} Phase diagrams under the annealed and quenched approaches for the $q$-voter model with $q=6$ and (a) independence (b) anticonformity. 
	Horizontal dashed lines are asymptotes that correspond to the results on a complete graph of the infinite size, i.e., the case when $\langle k\rangle\to\infty$.
	Panel~(c) magnifies the rectangular area marked in (a), whereas panels (d)-(f) illustrate the transitions at given values of the average node degree indicated in (c).
	Notation: $p$ -- the level of nonconformity,  $\langle k\rangle$ -- the average node degree, $c$ -- the concentration of voters with positive opinions, $q$ -- the influence group size.}
\end{figure*}

At the pair approximation level, the only network parameter that affects the model dynamics under the quenched approach is the average node degree.
The same applies to the $q$-voter models under the annealed approach \cite{Jed:17,Jed:etal:20,Gra:Kra:20,Vie:etal:20,Abr:Szn:20}.
Interestingly, if repetition of voters in the influence group is allowed, other moments of the degree distribution may appear in the solution \cite{Per:etal:18,Vie:etal:20}.
Figure~\ref{fig:phasemap} illustrates how the type of phase transition predicted by our approach depends on the average node degree, $\langle k\rangle$, and the influence group size, $q$.
However, these results should be treated with caution since the predictions of the pair approximation are not always correct in the case of the $q$-voter models with annealed disorder.
In the $q$-voter model with independence under the annealed approach,
the pair approximation predicts two consecutive phase transitions (one continuous and one discontinuous) for small enough average node degrees if $q>5$ [striped area in Fig.~\ref{fig:phasemap:a}; see also Fig.~\ref{fig:phasediagram:a}].
Nevertheless, instead of these two transitions, a single discontinuous phase transition shows up in the simulations \cite{Jed:17}.
On the other hand, in the $q$-voter model with anticonformity under the annealed approach, the pair approximation indicates that discontinuous phase transitions become continuous for large enough average node degrees [see Figs.~\ref{fig:phasemap:b} and \ref{fig:phasediagram:b}].
However, only continuous phase transitions have been observed in the simulations \cite{Jed:etal:20}.
For the quenched counterparts of the models, we have not observed similar errors. 
The pair approximation predicts only continuous phase transitions, and only such transitions have been detected in the simulations \cite{Szn:Szw:Wer:14,Jav:Squ:15}.
Therefore, the rounding effect actually occurs only in the $q$-voter model with independence.

\begin{figure}[!ht]
	\centering
	\subfloat{\label{fig:phases:a}}
	\subfloat{\label{fig:phases:b}}
	\epsfig{file=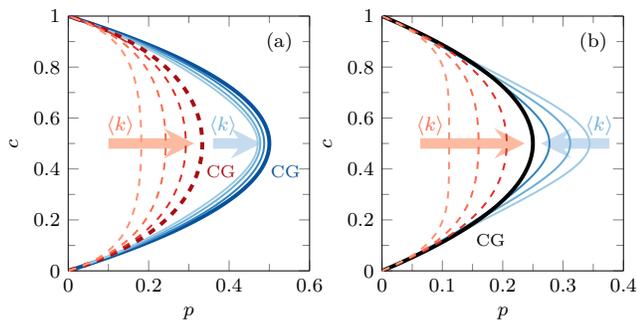,width=\linewidth}
	\caption{\label{fig:phases} Illustration how the average node degree of a network impacts phase diagrams under the annealed (dashed lines in shades of red) and quenched (continuous lines in shades of blue) approaches for the $q$-voter model with $q=2$ and (a) independence (b) anticonformity.
	Only stable concentrations that refer to the ordered phase are depicted for clarity.
    Three different values of the average node degree are considered, $\langle k\rangle\in\{4,6,12\}$. Arrows indicate the direction in which $\langle k\rangle$ increases. Thick lines (labeled by CG) correspond to the results on a complete graph of the infinite size, i.e., the case when $\langle k\rangle\to\infty$. Notation: $c$ -- the concentration of voters with positive opinions, $p$ -- the level of nonconformity,  $q$ -- the influence group size.}
\end{figure}

Let us now discuss how exactly the average node degree influences the phases. 
In the $q$-voter model with independence, the ordered phase becomes wider on networks with a larger average node degree regardless of the disorder type, see Figs.~\ref{fig:phasediagram:a} and \ref{fig:phases:a}.
For the annealed case, we can obtain the explicit formula for the transition point \cite{Jed:17}:
\begin{equation}
    p^*=\frac{q-1}{q-1+2^{q-1}\left(\frac{\langle k\rangle-1}{\langle k\rangle-2}\right)^q}.
\end{equation}
In the limit at infinity, $\langle k\rangle\to\infty$, this point tends to the one obtained for a well-mixed population given by the following formula \cite{Nyc:Szn:Cis:12,Nyc:Szn:13}:
\begin{equation}
p^*=\frac{q-1}{q-1+2^{q-1}}.   
\label{eq:tranpoint:ia}
\end{equation}
For the quenched case, the transition point is obtained numerically. In Fig.~\ref{fig:phasediagram:a}, we see that this point also tends to the one obtained for a well-mixed population \cite{Jed:Szn:17}, i.e.,
\begin{equation}
p^*=\frac{q-1}{q},
\end{equation}
as the average node degree becomes larger.
Note the change of disorder from annealed to quenched broadens the ordered phase. 
In Fig.~\ref{fig:phases:a}, we see that the quenched model is much less sensitive to the changes in $\langle k\rangle$.
Thus, it is easy to overlook the dependence of the transition point on the average node degree of a network when the model is studied based on simulations, as in Ref.~\cite{Szn:Szw:Wer:14}.

\begin{figure}[!t]
	\centering
	\subfloat{\label{fig:mc:a}}
	\subfloat{\label{fig:mc:b}}
	\subfloat{\label{fig:mc:c}}
	\subfloat{\label{fig:mc:d}}
	\epsfig{file=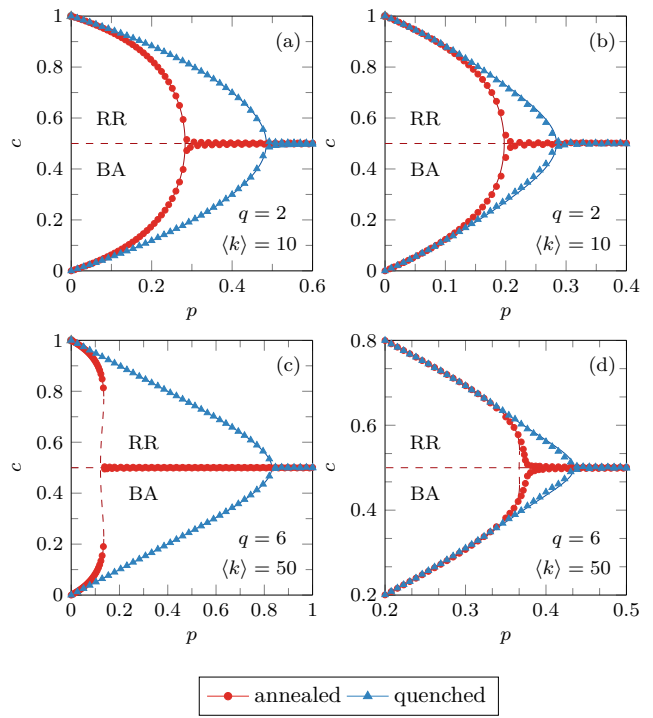,width=\linewidth}
	\caption{\label{fig:mc} Comparison between the annealed and quenched approaches for the $q$-voter model with (a, c) independence (b, d) anticonformity. 
	Lines refer to steady concentrations obtained form the pair approximation, whereas symbols refer to Monte Carlo simulations.
	The upper part of each diagram displays the simulation outcomes obtained on random regular graphs (RR), whereas the bottom part displays the simulation results obtained on Barab\'{a}si-Albert networks (BA).
	Notation: $c$ -- the concentration of voters with positive opinions, $p$ -- the level of nonconformity, $q$ -- the influence group size, $\langle k\rangle$ -- the average node degree.
	}
\end{figure}

In the $q$-voter model with anticonformity, the ordered phase becomes wider on networks with a larger average node degree only when annealed disorder is considered. In the presence of quenched disorder, the ordered phase shrinks as the average node degree becomes larger, see Figs.~\ref{fig:phasediagram:a} and \ref{fig:phases:a}.
For the annealed case, we can obtain the explicit formula for the transition point between the ordered and disordered phases \cite{Jed:etal:20}:
\begin{equation}
    p^*=\frac{q-1}{q-1+\left(1+q\frac{\langle k\rangle-2}{\langle k\rangle}\right)\left(\frac{\langle k\rangle}{\langle k\rangle -2}\right)^q}.
\end{equation}
For the quenched case, the transition point is obtained numerically.
In the limit at infinity, $\langle k\rangle\to\infty$, the transition points from both the approaches tend to the same point given by
\begin{equation}
    p^*=\frac{q-1}{2q},
\end{equation}
which is the point of the phase transition on a complete graph of an infinite size \cite{Jed:Szn:17}.
Thus, only in well-mixed populations, the annealed and quenched approaches lead to the same result.
On networks, however, the change of disorder form annealed to quenched alters the phase diagram in such a way that it broadens the ordered phase, just like in the $q$-voter model with independence but to a lesser extent [e.g., compare Figs.~\ref{fig:phasediagram:a} and \ref{fig:phasediagram:b}]. 
This can be also seen in Fig.~\ref{fig:mc} where the predictions of the pair approximations are presented together with the outcomes of Monte Carlo simulations.
Note that in Fig.~\ref{fig:mc:d}, the pair approximation incorrectly predicted the type of the phase transition for the $q$-voter model with anticonformity under the annealed approach.

\begin{figure}[!t]
	\centering
	\subfloat{\label{fig:partition:a}}
	\subfloat{\label{fig:partition:b}}
	\subfloat{\label{fig:partition:c}}
	\subfloat{\label{fig:partition:d}}
	\epsfig{file=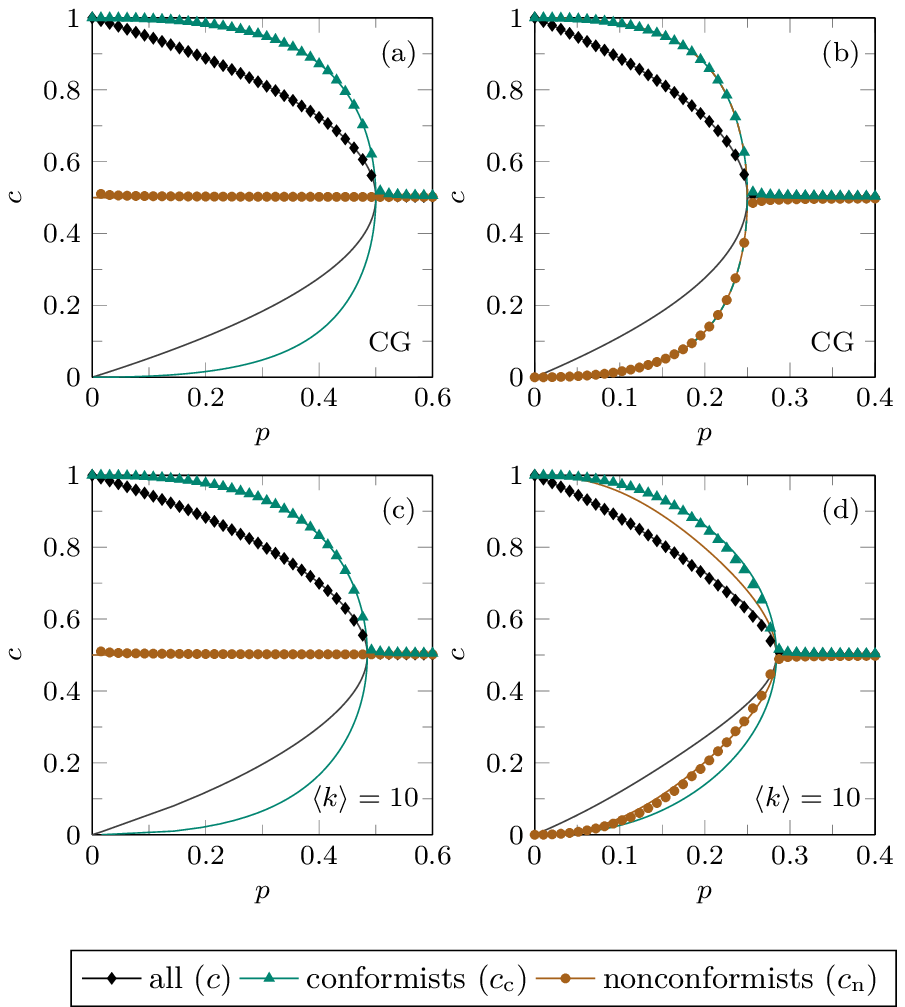,width=\linewidth}
	\caption{\label{fig:partition} Stable concentrations of voters with a positive opinion among different groups under the quenched approach [see Eq.~(\ref{eq:partition})] for the $q$-voter model with $q=2$ and (a, c) independence (b, d) anticonformity. The first row illustrates the results on a complete graph (CG), whereas the second one illustrates the results on a random regular graph with $\langle k\rangle=10$. Lines refer to the analytical predictions, whereas symbols refer to Monte Carlo simulations.
	Initially, all voters have a positive opinion in the simulations.
	Notation: $c$ -- the concentration of voters with positive opinions, $p$ -- the level of nonconformity, $\langle k\rangle$ -- the average node degree, $q$ -- the influence group size.}
\end{figure}

Lastly, under the quenched approach, we can study separately the concentration of positive agents among conformists and nonconformists, i.e., $c_\tau$ defined in Sec.~\ref{sec:pair}.
Figure~\ref{fig:partition} illustrates this partition.
In the $q$-voter model with independence, the situation on networks is similar to the one in a well-mixed population.
The concentration of positive agents among independent agents is always $1/2$ regardless of the average node degree, see Figs.~\ref{fig:partition:a} and \ref{fig:partition:c}.
On the other hand, the $q$-voter model with anticonformity behaves differently  on networks than on complete graphs.
For a well-mixed population, we have that $c_\text{c}=1-c_\text{a}$.
This means that there is the same level of agreement among agents in each of the groups, so the diagrams for conformists and anticonformists are on top of each other, see Fig.~\ref{fig:partition:b}.
However, this equality is no longer true on networks, where the agreement among conformists is higher than the one among anticonformists, so the diagram for conformists encloses the one for the anticonformists, see Fig.~\ref{fig:partition:d}.
Consequently, the network breaks the symmetry exhibited on a complete graph between the concentrations of conformists and anticonformists that hold the same opinion.

\section{Conclusions}
We examined how the change of disorder from annealed to quenched affects the phase diagrams exhibited by the $q$-voter model with independence and the $q$-voter model with anticonformity on networks.
To this end, we developed the pair approximation for the quenched versions of the models. 
This formalism can be also applied to other quenched dynamics of similar kind.
We validated our calculations by Monte Carlo simulations conducted on random regular graphs and Barab\'{a}si-Albert networks.

We showed that quenched disorder eliminates all discontinuous phase transitions in the $q$-voter model with independence (the $q$-voter model with anticofnormity does not exhibit discontinuous phase transitions at all).
Moreover, quenched disorder broadens the ordered phases of continuous phase transitions in both the models, but does so to a lesser extent in the $q$-voter model with anticonformity.
Similarly, quenched disorder broadens ordered phases of continuous phase transitions in the Galam model \cite{Gal:04,Sta:Sa:04}.

Our comparative analysis of the $q$-voter models with different types of nonconformity and disorder on networks revealed differences between them that cannot be observed in well-mixed populations.
Thus, relying only on the mean-field description, which does not take into account the network structure, may be misleading.
We showed that although annealed and quenched types of disorder lead to exactly the same result in the $q$-voter model with anticonformity at the mean-field level \cite{Jed:Szn:17}, these two disorder types do lead to distinct phase diagrams on networks.
These phase diagrams shift towards each other as the average node degree of a network increases, and eventually, they coincide in the limit of an infinite average node degree (i.e., the mean-field limit).
Therefore, the direction of a diagram shift depends on the disorder type in this case. 
For the annealed one, the phase diagrams shift towards higher values of the control parameter $p$, whereas for the
quenched one, they shift towards lower values of the control parameter $p$.
In contrast, for the $q$-voter model with independence, the phase diagrams move towards higher values of the control parameter no matter what the disorder type is, and they do not coincide even in the mean-field limit.

Moreover, there is a qualitative difference between the behavior of the $q$-voter model with anticonformity under the quenched approach on a network and on a complete graph.
On a network, the level of agreement among conformists is higher than the one among nonconformists, whereas in a well-mixed population, the level of agreement in both these groups is the same \cite{Jed:Szn:17}.

Finally, let us notice that the pair approximation may incorrectly predict the type of phase transitions displayed by the annealed models. 
Especially prone to this kind of error is the $q$-voter model with anticonformity.
Similar discrepancies are reported with regard to other dynamics \cite{Jed:etal:20,Vie:etal:20,Gra:Kra:20,Kra:20,Kra:21, Abr:etal:21}.
However, we have not observed such errors for the quenched counterparts of the studied models.
This demonstrates how accuracy of the pair approximation may change depending on the details of particular dynamics and cautions against relying solely on approximate methods.
On the other hand, well-conducted approximation may aid in discovering effects that are easy to overlook based on the simulations alone, as had been the case with the $q$-voter model with independence under the quenched approach. 
This particular model had been suggested to be insensitive to changes in the network structure \cite{Szn:Szw:Wer:14}.
However, with the support of the pair approximation, we were able to conclude that the network structure does influence all the $q$-voter models studied herein, including the $q$-voter model with independence under the quenched approach.

\section*{Acknowledgments}
This work was created as a result of Research Projects No. 2016/23/N/ST2/00729 (A.J.) and No. 2019/35/B/HS6/02530 (K.S.-W.) financed by the funds of the National Science Center (NCN, Poland), and it was supported in part by PLGrid Infrastructure.

\appendix
\section{DETAILS OF THE PAIR APPROXIMATION} 
\label{sec:apendix}
Herein, we present the details of the pair approximation for the $q$-voter models with nonconformity under the quenched approach (see Sec.~\ref{sec:pair}).
The calculations for the $q$-voter model with independence are included in first subsection, whereas those for the $q$-voter model with antifonformity are included in the second one.
In the following calculations, the average node degrees calculated only among nodes of type $\tau$ that are in state $j$ show up.
They are denoted by $\langle k_j^\tau\rangle$ and given by the following formulas:
\begin{align}
	\langle k_\uparrow^\text{c}\rangle=&\frac{e_{\uparrow\downarrow}^{\text{cc}}+e_{\uparrow\downarrow}^{\text{cn}}+e_{\uparrow\uparrow}^{\text{cc}}+e_{\uparrow\uparrow}^{\text{cn}}}{(1-p)c_\text{c}}\langle k\rangle,\\
	\langle k_\downarrow^\text{c}\rangle=&\frac{e_{\downarrow\uparrow}^{\text{cc}}+e_{\downarrow\uparrow}^{\text{cn}}+e_{\downarrow\downarrow}^{\text{cc}}+e_{\downarrow\downarrow}^{\text{cn}}}{(1-p)(1-c_\text{c})}\langle k\rangle,\\
	\langle k_\uparrow^\text{n}\rangle=&\frac{e_{\uparrow\downarrow}^{\text{nn}}+e_{\uparrow\downarrow}^{\text{nc}}+e_{\uparrow\uparrow}^{\text{nc}}+e_{\uparrow\uparrow}^{\text{nn}}}{pc_\text{n}}\langle k\rangle,\label{eq:avekui}\\
	\langle k_\downarrow^\text{n}\rangle=&\frac{e_{\downarrow\uparrow}^{\text{nn}}+e_{\downarrow\uparrow}^{\text{nc}}+e_{\downarrow\downarrow}^{\text{nn}}+e_{\downarrow\downarrow}^{\text{nc}}}{p(1-c_\text{n})}\langle k\rangle.\label{eq:avekdi}
\end{align}

\subsection{The $q$-voter model with independence} 
In this case, we only have conformists and independent individuals in the system, so $\tau\in\{\text{c},\text{i}\}$, and Eq.~(\ref{eq:dct}) gives:
\begin{align}
	\frac{dc_\text{c}}{dt}=&(1-c_\text{c})\bar{\alpha}^q-c_\text{c}\alpha^q,\\
	\frac{dc_\text{i}}{dt}=&\frac{1}{2}-c_\text{i}. \label{eq:dc_i}
\end{align}
From Eq.~(\ref{eq:dc_i}), we see right away that the steady concentration of independent individuals with a positive opinion is 
\begin{equation}
    {}_{\text{st}}c_\text{i}=1/2.
    \label{eq:stci}
\end{equation}
Note that we indicate the steady values by a lower index on the left hand side of a given quantity.
Below, we present the differential equations for the evolution of the concentrations of directed links obtain from Eq.~(\ref{eq:de}) after performing all the summations as in Ref.~\cite{Jed:17}.
Note that they include only the first moments of the degree distributions, i.e., $\langle k_j^\tau\rangle$.
Let us start with the concentrations of links that connect only conformists:
\begin{widetext}
\begin{align}
    \label{eq:deccuu}
	\frac{d}{dt} e_{\uparrow\uparrow}^{\text{cc}}=&\frac{2(1-p)}{\langle k\rangle}\left\lbrace (1-c_\text{c})\bar{\alpha}^q\left[q+\left(\langle k_\downarrow^\text{c}\rangle-q\right)\bar{\alpha}\right]\bar{\beta}-c_\text{c}\alpha^q\left(\langle k_\uparrow^\text{c}\rangle-q\right)(1-\alpha)\gamma\right\rbrace,\\
	\frac{d}{dt} e_{\downarrow\downarrow}^{\text{cc}}=&\frac{2(1-p)}{\langle k\rangle}\left\lbrace c_\text{c}\alpha^q\left[q+\left(\langle k_\uparrow^\text{c}\rangle-q\right)\alpha\right]\beta
	-(1-c_\text{c})\bar{\alpha}^q\left(\langle k_\downarrow^\text{c}\rangle-q\right)\left(1-\bar{\alpha}\right)\bar{\gamma}\right\rbrace,\\
	\frac{d}{dt} e_{\uparrow\downarrow}^{\text{cc}}=&\frac{1-p}{\langle k\rangle}c_\text{c}\alpha^q\left\lbrace\left(\langle k_\uparrow^\text{c}\rangle-q\right)\left(1-\alpha\right)\gamma
	-\left[q+\left(\langle k_\uparrow^\text{c}\rangle-q\right)\alpha\right]\beta\right\rbrace\nonumber\\
	&+\frac{1-p}{\langle k\rangle}(1-c_\text{c})\bar{\alpha}^q\left\lbrace\left(\langle k_\downarrow^\text{c}\rangle-q\right)\left(1-\bar{\alpha}\right)\bar{\gamma}
	-\left[q+\left(\langle k_\downarrow^\text{c}\rangle-q\right)\bar{\alpha}\right]\bar{\beta}\right\rbrace.
	\label{eq:deccud}
\end{align}
For the links that connect only independent individuals, we have the following:
\begin{align}
    	\frac{d}{dt} e_{\uparrow\uparrow}^{\text{ii}}=&\frac{p}{\langle k\rangle}\left\lbrace (1-c_\text{i})\langle k_\downarrow^\text{i}\rangle\widehat{\alpha}\widehat{\beta}-c_\text{i}\langle k_\uparrow^\text{i}\rangle\left(1-\widetilde{\alpha}\right)\widetilde{\gamma}\right\rbrace,\label{eq:deuuii}\\
	\frac{d}{dt} e_{\downarrow\downarrow}^{\text{ii}}=&\frac{p}{\langle k\rangle}\left\lbrace c_\text{i}\langle k_\uparrow^\text{i}\rangle\widetilde{\alpha}\widetilde{\beta}-(1-c_\text{i})\langle k_\downarrow^\text{i}\rangle\left(1-\widehat{\alpha}\right)\widehat{\gamma}\right\rbrace,\label{eq:deddii}\\
	\frac{d}{dt} e_{\uparrow\downarrow}^{\text{ii}}=&\frac{p}{2\langle k\rangle}c_\text{i}\left\lbrace \langle k_\uparrow^\text{i}\rangle\left(1-\widetilde{\alpha}\right)\widetilde{\gamma}-\langle k_\uparrow^\text{i}\rangle\widetilde{\alpha}\widetilde{\beta}\right\rbrace
	+\frac{p}{2\langle k\rangle}(1-c_\text{i})\left\lbrace \langle k_\downarrow^\text{i}\rangle\left(1-\widehat{\alpha}\right)\widehat{\gamma}-\langle k_\downarrow^\text{i}\rangle\widehat{\alpha}\widehat{\beta}\right\rbrace.
\end{align}
The last equations are for the concentrations of links that connect conformists and independent individuals. For those that link nodes in the same states, we have
\begin{align}
    \frac{d}{dt} e_{\uparrow\uparrow}^{\text{ci}}=&\frac{1-p}{\langle k\rangle}\left\lbrace(1-c_\text{c})\bar{\alpha}^q\left[q+\left(\langle k_\downarrow^\text{c}\rangle-q\right)\bar{\alpha}\right]\left(1-\bar{\beta}\right)
    -c_\text{c}\alpha^q\left(\langle k_\uparrow^\text{c}\rangle-q\right)(1-\alpha)(1-\gamma)\right\rbrace\nonumber\\
	&+\frac{p}{2\langle k\rangle}\left\lbrace(1-c_\text{i})\langle k_\downarrow^\text{i}\rangle\widehat{\alpha}\left(1-\widehat{\beta}\right)
	-c_\text{i}\langle k_\uparrow^\text{i}\rangle\left(1-\widetilde{\alpha}\right)\left(1-\widetilde{\gamma}\right)\right\rbrace,\\
	\frac{d}{dt} e_{\downarrow\downarrow}^{\text{ci}}=&\frac{1-p}{\langle k\rangle}\left\lbrace c_\text{c}\alpha^q\left[q+\left(\langle k_\uparrow^\text{c}\rangle-q\right)\alpha\right]\left(1-\beta\right)
	-(1-c_\text{c})\bar{\alpha}^q\left(\langle k_\downarrow^\text{c}\rangle-q\right)\left(1-\bar{\alpha}\right)\left(1-\bar{\gamma}\right)\right\rbrace\nonumber\\
	&+\frac{p}{2\langle k\rangle}\left\lbrace c_\text{i}\langle k_\uparrow^\text{i}\rangle\widetilde{\alpha}\left(1-\widetilde{\beta}\right)
	-(1-c_\text{i})\langle k_\downarrow^\text{i}\rangle\left(1-\widehat{\alpha}\right)\left(1-\widehat{\gamma}\right)\right\rbrace,
\end{align}
whereas for those that link nodes in different states, we have
\begin{align}
    \frac{d}{dt} e_{\uparrow\downarrow}^{\text{ci}}=&\frac{1-p}{\langle k\rangle}\left\lbrace (1-c_\text{c})\bar{\alpha}^q\left(\langle k_\downarrow^\text{c}\rangle-q\right)\left(1-\bar{\alpha}\right)\left(1-\bar{\gamma}\right)
    -c_\text{c}\alpha^q\left[q+\left(\langle k_\uparrow^\text{c}\rangle-q\right)\alpha\right]\left(1-\beta\right)\right\rbrace\nonumber\\
	&+\frac{p}{2\langle k\rangle}\left\lbrace c_\text{i}\langle k_\uparrow^\text{i}\rangle\left(1-\widetilde{\alpha}\right)\left(1-\widetilde{\gamma}\right)
	-(1-c_\text{i})\langle k_\downarrow^\text{i}\rangle\widehat{\alpha}\left(1-\widehat{\beta}\right)\right\rbrace,\\
	\frac{d}{dt} e_{\downarrow\uparrow}^{\text{ci}}=&\frac{1-p}{\langle k\rangle}\left\lbrace c_\text{c}\alpha^q\left(\langle k_\uparrow^\text{c}\rangle-q\right)\left(1-\alpha\right)\left(1-\gamma\right)
	-(1-c_\text{c})\bar{\alpha}^q\left[q+\left(\langle k_\downarrow^\text{c}\rangle-q\right)\bar{\alpha}\right]\left(1-\bar{\beta}\right)\right\rbrace\nonumber\\
	&+\frac{p}{2\langle k\rangle}\left\lbrace (1-c_\text{i})\langle k_\downarrow^\text{i}\rangle\left(1-\widehat{\alpha}\right)\left(1-\widehat{\gamma}\right)
	-c_\text{i}\langle k_\uparrow^\text{i}\rangle\widetilde{\alpha}\left(1-\widetilde{\beta}\right)\right\rbrace.
\end{align}
\end{widetext}
Having the above equations, we can write down the differential equations that set the time evolution of the average node degrees defined at the beginning of Appendix~\ref{sec:apendix}. They are given by the following:
\begin{align}
	\frac{d}{dt}\langle k_\uparrow^\text{c}\rangle=&\frac{1-c_\text{c}}{c_\text{c}}\bar{\alpha}^q\left[\langle k_\downarrow^\text{c}\rangle-\langle k_\uparrow^\text{c}\rangle\right],\label{eq:dkcup}\\
	\frac{d}{dt}\langle k_\downarrow^\text{c}\rangle=&\frac{c_\text{c}}{1-c_\text{c}}\alpha^q\left[\langle k_\uparrow^\text{c}\rangle-\langle k_\downarrow^\text{c}\rangle\right],\\
	\frac{d}{dt}\langle k_\uparrow^\text{i}\rangle=&\frac{1-c_\text{i}}{2c_\text{i}}\left[\langle k_\downarrow^\text{i}\rangle-\langle k_\uparrow^\text{i}\rangle\right],\\
	\frac{d}{dt}\langle k_\downarrow^\text{i}\rangle=&\frac{c_\text{i}}{2(1-c_\text{i})}\left[\langle k_\uparrow^\text{i}\rangle-\langle k_\downarrow^\text{i}\rangle\right].\label{eq:dkidown}
\end{align}
Thus, in the steady state, we have that
\begin{equation}
    {}_{\text{st}}\langle k_\uparrow^\text{c}\rangle={}_{\text{st}}\langle k_\downarrow^\text{c}\rangle,
\end{equation} 
and
\begin{equation}
   {}_{\text{st}}\langle k_\uparrow^\text{i}\rangle={}_{\text{st}}\langle k_\downarrow^\text{i}\rangle.
    \label{eq:avedegrees}
\end{equation} 
The steady solutions of the concentrations of links that connect independent individuals can be found by combining Eqs.~(\ref{eq:deuuii})  and (\ref{eq:deddii}) together with Eqs.~(\ref{eq:sumeii}), (\ref{eq:stci}), (\ref{eq:avedegrees}), and the definitions of the average node degrees, i.e., Eqs.~(\ref{eq:avekui}) and (\ref{eq:avekdi}):
\begin{equation}
    {}_{\text{st}}e_{\uparrow\uparrow}^{\text{ii}}={}_{\text{st}}e_{\downarrow\downarrow}^{\text{ii}}={}_{\text{st}}e_{\uparrow\downarrow}^{\text{ii}}=\frac{1}{4}p^2.
\end{equation}
The steady solutions for the rest of the concentrations are found numerically after some simplifications of the equations.

\subsection{The $q$-voter model with anticonformity}  
In this case, we only have conformists and anticonformists in the system, so $\tau\in\{\text{c},\text{a}\}$, and Eq.~(\ref{eq:dct}) gives:
\begin{align}
	\frac{dc_\text{c}}{dt}=&(1-c_\text{c})\bar{\alpha}^q-c_\text{c}\alpha^q,\\
	\frac{dc_\text{a}}{dt}=&(1-c_\text{a})\left(1-\widehat{\alpha}\right)^q-c_\text{a}\left(1-\widetilde{\alpha}\right)^q.
\end{align}
Below, we present the differential equations for the evolution of the concentrations of directed links obtain from Eq.~(\ref{eq:de}) after performing all the summations as in Ref.~\cite{Jed:17}.
Note that they include only the first moments of the degree distributions, i.e., $\langle k_j^\tau\rangle$.
The equations for the concentrations of the links that connect only conformists are the same as in the case of the $q$-voter model with independence, so they are given by Eqs.~(\ref{eq:deccuu})-(\ref{eq:deccud}).
On the other hand, the differential equations for the links that connect only anticonformists are the following:
\begin{widetext}
\begin{align}
    \frac{d}{dt} e_{\uparrow\uparrow}^{\text{aa}}=&\frac{2p}{\langle k\rangle}\left\lbrace (1-c_\text{a})\left(1-\widehat{\alpha}\right)^q\left(\langle k_\downarrow^\text{a}\rangle-q\right)\widehat{\alpha}\widehat{\beta}
    -c_\text{a}\left(1-\widetilde{\alpha}\right)^q\left[q+\left(\langle k_\uparrow^\text{a}\rangle-q\right)\left(1-\widetilde{\alpha}\right)\right]\widetilde{\gamma}\right\rbrace,\\
	\frac{d}{dt} e_{\downarrow\downarrow}^{\text{aa}}=&\frac{2p}{\langle k\rangle}\left\lbrace c_\text{a}\left(1-\widetilde{\alpha}\right)^q\left(\langle k_\uparrow^\text{a}\rangle-q\right)\widetilde{\alpha}\widetilde{\beta}
	-(1-c_\text{a})\left(1-\widehat{\alpha}\right)^q\left[q+\left(\langle k_\downarrow^a\rangle-q\right)\left(1-\widehat{\alpha}\right)\right]\widehat{\gamma}\right\rbrace,\\
	\frac{d}{dt} e_{\uparrow\downarrow}^{\text{aa}}=&\frac{p}{\langle k\rangle}c_\text{a}\left(1-\widetilde{\alpha}\right)^q\left\lbrace \left[q+\left(\langle k_\uparrow^\text{a}\rangle-q\right)\left(1-\widetilde{\alpha}\right)\right]\widetilde{\gamma}
	-\left(\langle k_\uparrow^\text{a}\rangle-q\right)\widetilde{\alpha}\widetilde{\beta}\right\rbrace\nonumber\\
	&+\frac{p}{\langle k\rangle}(1-c_\text{a})\left(1-\widehat{\alpha}\right)^q\left\lbrace \left[q+\left(\langle k_\downarrow^\text{a}\rangle-q\right)\left(1-\widehat{\alpha}\right)\right]\widehat{\gamma}
	-\left(\langle k_\downarrow^\text{a}\rangle-q\right)\widehat{\alpha}\widehat{\beta}\right\rbrace.
\end{align}
The last equations are for the concentrations of links that connect conformists and anticonformists. For those that link nodes in the same states, we have
\begin{align}
    \frac{d}{dt} e_{\uparrow\uparrow}^{\text{ca}}=&\frac{1-p}{\langle k\rangle}\left\lbrace (1-c_\text{c})\bar{\alpha}^q\left[q+\left(\langle k_\downarrow^\text{c}\rangle-q\right)\bar{\alpha}\right](1-\bar{\beta})
    -c_\text{c}\alpha^q\left(\langle k_\uparrow^\text{c}\rangle-q\right)\left(1-\alpha\right)\left(1-\gamma\right)\right\rbrace\nonumber\\
	&+\frac{p}{\langle k\rangle}\left\lbrace (1-c_\text{a})\left(1-\widehat{\alpha}\right)^q\left(\langle k_\downarrow^\text{a}\rangle-q\right)\widehat{\alpha}\left(1-\widehat{\beta}\right)
	-c_\text{a}\left(1-\widetilde{\alpha}\right)^q\left[q+\left(\langle k_\uparrow^\text{a}\rangle-q\right)\left(1-\widetilde{\alpha}\right)\right]\left(1-\widetilde{\gamma}\right)\right\rbrace,\\
	\frac{d}{dt} e_{\downarrow\downarrow}^{\text{ca}}=&\frac{1-p}{\langle k\rangle}\left\lbrace c_\text{c}\alpha^q\left[q+\left(\langle k_\uparrow^\text{c}\rangle-q\right)\alpha\right]\left(1-\beta\right)
	-(1-c_\text{c})\bar{\alpha}^q\left(\langle k_\downarrow^\text{c}\rangle-q\right)\left(1-\bar{\alpha}\right)\left(1-\bar{\gamma}\right)\right\rbrace\nonumber\\
	&+\frac{p}{\langle k\rangle}\left\lbrace c_\text{a}\left(1-\widetilde{\alpha}\right)^q\left(\langle k_\uparrow^\text{a}\rangle-q\right)\widetilde{\alpha}\left(1-\widetilde{\beta}\right)
	-(1-c_\text{a})\left(1-\widehat{\alpha}\right)^q\left[q+\left(\langle k_\downarrow^a\rangle-q\right)\left(1-\widehat{\alpha}\right)\right]\left(1-\widehat{\gamma}\right)\right\rbrace,
\end{align}
whereas for those that link nodes in different states, we have
\begin{align}
    \frac{d}{dt} e_{\uparrow\downarrow}^{\text{ca}}=&\frac{1-p}{\langle k\rangle}\left\lbrace (1-c_\text{c})\bar{\alpha}^q\left(\langle k_\downarrow^\text{c}\rangle-q\right)\left(1-\bar{\alpha}\right)\left(1-\bar{\gamma}\right)
    -c_\text{c}\alpha^q\left[q+\left(\langle k_\uparrow^\text{c}\rangle-q\right)\alpha\right]\left(1-\beta\right)\right\rbrace\nonumber\\
	&+\frac{p}{\langle k\rangle}\left\lbrace c_\text{a}\left(1-\widetilde{\alpha}\right)^q\left[q+\left(\langle k_\uparrow^\text{a}\rangle-q\right)\left(1-\widetilde{\alpha}\right)\right]\left(1-\widetilde{\gamma}\right)
	-(1-c_\text{a})\left(1-\widehat{\alpha}\right)^q\left(\langle k_\downarrow^\text{a}\rangle-q\right)\widehat{\alpha}\left(1-\widehat{\beta}\right)\right\rbrace,\\
	\frac{d}{dt} e_{\downarrow\uparrow}^{\text{ca}}=&\frac{1-p}{\langle k\rangle}\left\lbrace c_\text{c}\alpha^q\left(\langle k_\uparrow^\text{c}\rangle-q\right)\left(1-\alpha\right)\left(1-\gamma\right)
	-(1-c_\text{c})\bar{\alpha}^q\left[q+\left(\langle k_\downarrow^\text{c}\rangle-q\right)\bar{\alpha}\right]\left(1-\bar{\beta}\right)\right\rbrace\nonumber\\
	&+\frac{p}{\langle k\rangle}\left\lbrace (1-c_\text{a})\left(1-\widehat{\alpha}\right)^q\left[q+\left(\langle k_\downarrow^\text{a}\rangle-q\right)\left(1-\widehat{\alpha}\right)\right]\left(1-\widehat{\gamma}\right)
	-c_\text{a}\left(1-\widetilde{\alpha}\right)^q\left(\langle k_\uparrow^\text{a}\rangle-q\right)\widetilde{\alpha}\left(1-\widetilde{\beta}\right)\right\rbrace.
\end{align}
\end{widetext}
Having the above equations, we can write down the differential equations that set the time evolution of the average node degrees defined at the beginning of Appendix~\ref{sec:apendix}. They are given by the following:
\begin{align}
	\frac{d}{dt}\langle k_\uparrow^\text{c}\rangle=&\frac{1-c_\text{c}}{c_\text{c}}\bar{\alpha}^q\left[\langle k_\downarrow^\text{c}\rangle-\langle k_\uparrow^c\rangle\right],\label{eq:dkcupant}\\
	\frac{d}{dt}\langle k_\downarrow^\text{c}\rangle=&\frac{c_\text{c}}{1-c_\text{c}}\alpha^q\left[\langle k_\uparrow^\text{c}\rangle-\langle k_\downarrow^\text{c}\rangle\right],\\
	\frac{d}{dt}\langle k_\uparrow^\text{a}\rangle=&\frac{1-c_\text{a}}{c_\text{a}}\left(1-\widehat{\alpha}\right)^q\left[\langle k_\downarrow^\text{a}\rangle-\langle k_\uparrow^\text{a}\rangle\right],\\
	\frac{d}{dt}\langle k_\downarrow^\text{a}\rangle=&\frac{c_\text{a}}{1-c_\text{a}}\left(1-\widetilde{\alpha}\right)^q\left[\langle k_\uparrow^\text{a}\rangle-\langle k_\downarrow^\text{a}\rangle\right].\label{eq:dkidownant}
\end{align}
Thus, in the steady state, we have that
\begin{equation}
    {}_{\text{st}}\langle k_\uparrow^\text{c}\rangle={}_{\text{st}}\langle k_\downarrow^\text{c}\rangle,
    \label{eq:avedegreesant}
\end{equation} 
and
\begin{equation}
    {}_{\text{st}}\langle k_\uparrow^\text{a}\rangle={}_{\text{st}}\langle k_\downarrow^\text{a}\rangle.
    \label{eq:avedegreesant}
\end{equation} 
All the steady solutions for the concentrations are found numerically after some simplifications of the equations.

\section{DETAILS OF MONTE CARLO SIMULATIONS} 
\label{sec:apendix-b}
We carry out Monte Carlo simulations to validate our analytical calculations.
The simulations are performed on random regular graphs \cite{Ruc:Wor:92} and Barab\'{a}si-Albert networks \cite{Bar:Alb:99}.
The former are characterized by the same number of neighbors for all the nodes, so in this sense such networks are homogeneous structures.
In contrast, the later belong to the class of scale-free networks \cite{Boc:etal:06}, which are characterized by the power law tail in their degree distribution, i.e., $P(k)\sim k^{-\lambda}$. 
This means that some nodes of the network have many more neighbors than others, which is a common feature for many real-life structures \cite{Boc:etal:06}.
For Barab\'{a}si-Albert networks, the tail exponent is $\lambda=3$.

To construct a random regular graph, we start with $N$ isolated nodes, and we add edges one by one by connecting randomly pairs of nodes so that the maximum degree of the network does not exceed a desired number \cite{Ruc:Wor:92}.
If the resulted graph is regular, we accept it, otherwise we repeat the construction process.
On the other hand, a Barab\'{a}si-Albert network is created using the preferential attachment mechanism  \cite{Bar:Alb:99}.
We start with a small number of fully connected nodes. New nodes are added to the network one by one until the resulting structure has $N$ nodes. Each new node is connected to $n$ already existing nodes selected with probability proportional to the number of neighbors that the nodes already have.
The average node degree of such a network is $\langle k\rangle=2n$.

One Monte Carlo step (MCS) is understood as $N$ updates of the model described in Sec.~\ref{sec:model}.
In the simulations, we measure the absolute value of the average opinion (an analog to the magnetization per spin):
\begin{equation}
    m=\left|\frac{1}{N}\sum_i^N o_i\right|,
    \label{eq:m}
\end{equation}
where $o_i$ is the opinion of the $i$-th agent in the system.
To construct the phase diagrams, we consider the mean value of $m$ denoted by $M=\left[\langle m\rangle_t\right]_s$.
The angle brackets, $\langle \cdot\rangle_t$, represent the average performed over time in the stationary regime during a single simulation.
We skipped the first $4000$~MCS to let the system reach the stationary state and
performed the time average  over next $2000$~MCS.
The square brackets, $[ \cdot]_s$, represent the sample average that was performed over 10 independent simulations with different realizations of the networks.
Due to big system sizes and small fluctuations for considered values of $p$, this is enough to obtain small statistical errors (of the mark size order).

The concentration of agents with positive opinions presented in Figs.~\ref{fig:mc} and \ref{fig:partition} corresponds to $\frac{1}{2}\big(1+M\big)$ and $\frac{1}{2}\big(1-M\big)$ for the upper and the lower part of the diagrams, respectively. 
Simulated systems include $N=10^5$ voters.

\bibliography{02092020_literature}

\begin{thebibliography}{50}%
\makeatletter
\providecommand \@ifxundefined [1]{%
 \@ifx{#1\undefined}
}%
\providecommand \@ifnum [1]{%
 \ifnum #1\expandafter \@firstoftwo
 \else \expandafter \@secondoftwo
 \fi
}%
\providecommand \@ifx [1]{%
 \ifx #1\expandafter \@firstoftwo
 \else \expandafter \@secondoftwo
 \fi
}%
\providecommand \natexlab [1]{#1}%
\providecommand \enquote  [1]{``#1''}%
\providecommand \bibnamefont  [1]{#1}%
\providecommand \bibfnamefont [1]{#1}%
\providecommand \citenamefont [1]{#1}%
\providecommand \href@noop [0]{\@secondoftwo}%
\providecommand \href [0]{\begingroup \@sanitize@url \@href}%
\providecommand \@href[1]{\@@startlink{#1}\@@href}%
\providecommand \@@href[1]{\endgroup#1\@@endlink}%
\providecommand \@sanitize@url [0]{\catcode `\\12\catcode `\$12\catcode
  `\&12\catcode `\#12\catcode `\^12\catcode `\_12\catcode `\%12\relax}%
\providecommand \@@startlink[1]{}%
\providecommand \@@endlink[0]{}%
\providecommand \url  [0]{\begingroup\@sanitize@url \@url }%
\providecommand \@url [1]{\endgroup\@href {#1}{\urlprefix }}%
\providecommand \urlprefix  [0]{URL }%
\providecommand \Eprint [0]{\href }%
\providecommand \doibase [0]{http://dx.doi.org/}%
\providecommand \selectlanguage [0]{\@gobble}%
\providecommand \bibinfo  [0]{\@secondoftwo}%
\providecommand \bibfield  [0]{\@secondoftwo}%
\providecommand \translation [1]{[#1]}%
\providecommand \BibitemOpen [0]{}%
\providecommand \bibitemStop [0]{}%
\providecommand \bibitemNoStop [0]{.\EOS\space}%
\providecommand \EOS [0]{\spacefactor3000\relax}%
\providecommand \BibitemShut  [1]{\csname bibitem#1\endcsname}%
\let\auto@bib@innerbib\@empty
\bibitem [{\citenamefont {Dorogovtsev}\ \emph {et~al.}(2008)\citenamefont
  {Dorogovtsev}, \citenamefont {Goltsev},\ and\ \citenamefont
  {Mendes}}]{Dor:Gol:Men:08}%
  \BibitemOpen
  \bibfield  {author} {\bibinfo {author} {\bibfnamefont {S.~N.}\ \bibnamefont
  {Dorogovtsev}}, \bibinfo {author} {\bibfnamefont {A.~V.}\ \bibnamefont
  {Goltsev}}, \ and\ \bibinfo {author} {\bibfnamefont {J.~F.~F.}\ \bibnamefont
  {Mendes}},\ }\bibfield  {title} {\enquote {\bibinfo {title} {Critical
  phenomena in complex networks},}\ }\href {\doibase
  10.1103/RevModPhys.80.1275} {\bibfield  {journal} {\bibinfo  {journal} {Rev.
  Mod. Phys.}\ }\textbf {\bibinfo {volume} {80}},\ \bibinfo {pages}
  {1275--1335} (\bibinfo {year} {2008})}\BibitemShut {NoStop}%
\bibitem [{\citenamefont {Castellano}\ \emph
  {et~al.}(2009{\natexlab{a}})\citenamefont {Castellano}, \citenamefont
  {Fortunato},\ and\ \citenamefont {Loreto}}]{Cas:For:Lor:09}%
  \BibitemOpen
  \bibfield  {author} {\bibinfo {author} {\bibfnamefont {C.}~\bibnamefont
  {Castellano}}, \bibinfo {author} {\bibfnamefont {S.}~\bibnamefont
  {Fortunato}}, \ and\ \bibinfo {author} {\bibfnamefont {V.}~\bibnamefont
  {Loreto}},\ }\bibfield  {title} {\enquote {\bibinfo {title} {Statistical
  physics of social dynamics},}\ }\href {\doibase 10.1103/RevModPhys.81.591}
  {\bibfield  {journal} {\bibinfo  {journal} {Rev. Mod. Phys.}\ }\textbf
  {\bibinfo {volume} {81}},\ \bibinfo {pages} {591--646} (\bibinfo {year}
  {2009}{\natexlab{a}})}\BibitemShut {NoStop}%
\bibitem [{\citenamefont {Mu\~noz}\ \emph {et~al.}(2010)\citenamefont
  {Mu\~noz}, \citenamefont {Juh\'asz}, \citenamefont {Castellano},\ and\
  \citenamefont {\'Odor}}]{Mun:etal:10}%
  \BibitemOpen
  \bibfield  {author} {\bibinfo {author} {\bibfnamefont {M.~A.}\ \bibnamefont
  {Mu\~noz}}, \bibinfo {author} {\bibfnamefont {R.}~\bibnamefont {Juh\'asz}},
  \bibinfo {author} {\bibfnamefont {C.}~\bibnamefont {Castellano}}, \ and\
  \bibinfo {author} {\bibfnamefont {G.}~\bibnamefont {\'Odor}},\ }\bibfield
  {title} {\enquote {\bibinfo {title} {Griffiths phases on complex networks},}\
  }\href {\doibase 10.1103/PhysRevLett.105.128701} {\bibfield  {journal}
  {\bibinfo  {journal} {Phys. Rev. Lett.}\ }\textbf {\bibinfo {volume} {105}},\
  \bibinfo {pages} {128701} (\bibinfo {year} {2010})}\BibitemShut {NoStop}%
\bibitem [{\citenamefont {Mancastroppa}\ \emph {et~al.}(2021)\citenamefont
  {Mancastroppa}, \citenamefont {Castellano}, \citenamefont {Vezzani},\ and\
  \citenamefont {Burioni}}]{Man:etal:21}%
  \BibitemOpen
  \bibfield  {author} {\bibinfo {author} {\bibfnamefont {M.}~\bibnamefont
  {Mancastroppa}}, \bibinfo {author} {\bibfnamefont {C.}~\bibnamefont
  {Castellano}}, \bibinfo {author} {\bibfnamefont {A.}~\bibnamefont {Vezzani}},
  \ and\ \bibinfo {author} {\bibfnamefont {R.}~\bibnamefont {Burioni}},\
  }\bibfield  {title} {\enquote {\bibinfo {title} {Stochastic sampling effects
  favor manual over digital contact tracing},}\ }\href {\doibase
  10.1038/s41467-021-22082-7} {\bibfield  {journal} {\bibinfo  {journal} {Nat.
  Commun.}\ }\textbf {\bibinfo {volume} {12}},\ \bibinfo {pages} {1919}
  (\bibinfo {year} {2021})}\BibitemShut {NoStop}%
\bibitem [{\citenamefont {\'Odor}\ and\ \citenamefont
  {de~Simoni}(2021)}]{Odo:Sim:21}%
  \BibitemOpen
  \bibfield  {author} {\bibinfo {author} {\bibfnamefont {G.}~\bibnamefont
  {\'Odor}}\ and\ \bibinfo {author} {\bibfnamefont {B.}~\bibnamefont
  {de~Simoni}},\ }\bibfield  {title} {\enquote {\bibinfo {title} {Heterogeneous
  excitable systems exhibit griffiths phases below hybrid phase transitions},}\
  }\href {\doibase 10.1103/PhysRevResearch.3.013106} {\bibfield  {journal}
  {\bibinfo  {journal} {Phys. Rev. Research}\ }\textbf {\bibinfo {volume}
  {3}},\ \bibinfo {pages} {013106} (\bibinfo {year} {2021})}\BibitemShut
  {NoStop}%
\bibitem [{\citenamefont {Sznajd-Weron}\ \emph {et~al.}(2014)\citenamefont
  {Sznajd-Weron}, \citenamefont {Szwabiński},\ and\ \citenamefont
  {Weron}}]{Szn:Szw:Wer:14}%
  \BibitemOpen
  \bibfield  {author} {\bibinfo {author} {\bibfnamefont {K.}~\bibnamefont
  {Sznajd-Weron}}, \bibinfo {author} {\bibfnamefont {J.}~\bibnamefont
  {Szwabiński}}, \ and\ \bibinfo {author} {\bibfnamefont {R.}~\bibnamefont
  {Weron}},\ }\bibfield  {title} {\enquote {\bibinfo {title} {Is the
  person-situation debate important for agent-based modeling and vice-versa?}}\
  }\href {\doibase 10.1371/journal.pone.0112203} {\bibfield  {journal}
  {\bibinfo  {journal} {PLoS ONE}\ }\textbf {\bibinfo {volume} {9}},\ \bibinfo
  {pages} {e112203} (\bibinfo {year} {2014})}\BibitemShut {NoStop}%
\bibitem [{\citenamefont {Cheon}\ and\ \citenamefont
  {Galam}(2018)}]{Che:Gal:18}%
  \BibitemOpen
  \bibfield  {author} {\bibinfo {author} {\bibfnamefont {T.}~\bibnamefont
  {Cheon}}\ and\ \bibinfo {author} {\bibfnamefont {S.}~\bibnamefont {Galam}},\
  }\bibfield  {title} {\enquote {\bibinfo {title} {Dynamical {G}alam model},}\
  }\href {\doibase 10.1016/j.physleta.2018.04.019} {\bibfield  {journal}
  {\bibinfo  {journal} {Phys. Lett. A}\ }\textbf {\bibinfo {volume} {382}},\
  \bibinfo {pages} {1509--1515} (\bibinfo {year} {2018})}\BibitemShut {NoStop}%
\bibitem [{\citenamefont {J\k{e}drzejewski}\ and\ \citenamefont
  {Sznajd-Weron}(2019)}]{Jed:Szn:19}%
  \BibitemOpen
  \bibfield  {author} {\bibinfo {author} {\bibfnamefont {A.}~\bibnamefont
  {J\k{e}drzejewski}}\ and\ \bibinfo {author} {\bibfnamefont {K.}~\bibnamefont
  {Sznajd-Weron}},\ }\bibfield  {title} {\enquote {\bibinfo {title}
  {Statistical {P}hysics {O}f {O}pinion {F}ormation: Is it a {SPOOF}?}}\ }\href
  {\doibase https://doi.org/10.1016/j.crhy.2019.05.002} {\bibfield  {journal}
  {\bibinfo  {journal} {C. R. Physique}\ }\textbf {\bibinfo {volume} {20}},\
  \bibinfo {pages} {244--261} (\bibinfo {year} {2019})}\BibitemShut {NoStop}%
\bibitem [{\citenamefont {Aizenman}\ and\ \citenamefont
  {Wehr}(1990)}]{Aiz:Weh:90}%
  \BibitemOpen
  \bibfield  {author} {\bibinfo {author} {\bibfnamefont {M.}~\bibnamefont
  {Aizenman}}\ and\ \bibinfo {author} {\bibfnamefont {J.}~\bibnamefont
  {Wehr}},\ }\bibfield  {title} {\enquote {\bibinfo {title} {Rounding effects
  of quenched randomness on first-order phase transitions},}\ }\href {\doibase
  10.1007/BF02096933} {\bibfield  {journal} {\bibinfo  {journal} {Commun. Math.
  Phys.}\ }\textbf {\bibinfo {volume} {130}},\ \bibinfo {pages} {489--528}
  (\bibinfo {year} {1990})}\BibitemShut {NoStop}%
\bibitem [{\citenamefont {Borile}\ \emph {et~al.}(2013)\citenamefont {Borile},
  \citenamefont {Maritan},\ and\ \citenamefont {Mu\~noz}}]{Bor:Mar:Mun:13}%
  \BibitemOpen
  \bibfield  {author} {\bibinfo {author} {\bibfnamefont {C.}~\bibnamefont
  {Borile}}, \bibinfo {author} {\bibfnamefont {A.}~\bibnamefont {Maritan}}, \
  and\ \bibinfo {author} {\bibfnamefont {M.~A.}\ \bibnamefont {Mu\~noz}},\
  }\bibfield  {title} {\enquote {\bibinfo {title} {The effect of quenched
  disorder in neutral theories},}\ }\href {\doibase
  10.1088/1742-5468/2013/04/P04032} {\bibfield  {journal} {\bibinfo  {journal}
  {J. Stat. Mech.: Theory Exp.}\ }\textbf {\bibinfo {volume} {2013}},\ \bibinfo
  {pages} {P04032} (\bibinfo {year} {2013})}\BibitemShut {NoStop}%
\bibitem [{\citenamefont {Villa~Mart\'{\i}n}\ \emph {et~al.}(2014)\citenamefont
  {Villa~Mart\'{\i}n}, \citenamefont {Bonachela},\ and\ \citenamefont
  {Mu\~noz}}]{Vil:Bon:Mun:14}%
  \BibitemOpen
  \bibfield  {author} {\bibinfo {author} {\bibfnamefont {P.}~\bibnamefont
  {Villa~Mart\'{\i}n}}, \bibinfo {author} {\bibfnamefont {J.~A.}\ \bibnamefont
  {Bonachela}}, \ and\ \bibinfo {author} {\bibfnamefont {M.~A.}\ \bibnamefont
  {Mu\~noz}},\ }\bibfield  {title} {\enquote {\bibinfo {title} {Quenched
  disorder forbids discontinuous transitions in nonequilibrium low-dimensional
  systems},}\ }\href {\doibase 10.1103/PhysRevE.89.012145} {\bibfield
  {journal} {\bibinfo  {journal} {Phys. Rev. E}\ }\textbf {\bibinfo {volume}
  {89}},\ \bibinfo {pages} {012145} (\bibinfo {year} {2014})}\BibitemShut
  {NoStop}%
\bibitem [{\citenamefont {J\k{e}drzejewski}\ and\ \citenamefont
  {Sznajd-Weron}(2020)}]{Jed:Szn:20}%
  \BibitemOpen
  \bibfield  {author} {\bibinfo {author} {\bibfnamefont {A.}~\bibnamefont
  {J\k{e}drzejewski}}\ and\ \bibinfo {author} {\bibfnamefont {K.}~\bibnamefont
  {Sznajd-Weron}},\ }\bibfield  {title} {\enquote {\bibinfo {title} {Nonlinear
  $q$-voter model from the quenched perspective},}\ }\href {\doibase
  10.1063/1.5134684} {\bibfield  {journal} {\bibinfo  {journal} {Chaos}\
  }\textbf {\bibinfo {volume} {30}},\ \bibinfo {pages} {013150} (\bibinfo
  {year} {2020})}\BibitemShut {NoStop}%
\bibitem [{\citenamefont {Nowak}\ \emph {et~al.}(2021)\citenamefont {Nowak},
  \citenamefont {Sto{\'n}},\ and\ \citenamefont
  {Sznajd-Weron}}]{Now:Sto:Szn:21}%
  \BibitemOpen
  \bibfield  {author} {\bibinfo {author} {\bibfnamefont {B.}~\bibnamefont
  {Nowak}}, \bibinfo {author} {\bibfnamefont {B.}~\bibnamefont {Sto{\'n}}}, \
  and\ \bibinfo {author} {\bibfnamefont {K.}~\bibnamefont {Sznajd-Weron}},\
  }\bibfield  {title} {\enquote {\bibinfo {title} {Discontinuous phase
  transitions in the multi-state noisy $q$-voter model: quenched vs. annealed
  disorder},}\ }\href {\doibase 10.1038/s41598-021-85361-9} {\bibfield
  {journal} {\bibinfo  {journal} {Sci. Rep.}\ }\textbf {\bibinfo {volume}
  {11}},\ \bibinfo {pages} {1--13} (\bibinfo {year} {2021})}\BibitemShut
  {NoStop}%
\bibitem [{\citenamefont {Nowak}\ and\ \citenamefont
  {Sznajd-Weron}(2022)}]{Now:Szn:21}%
  \BibitemOpen
  \bibfield  {author} {\bibinfo {author} {\bibfnamefont {Bart\l{}omiej}\
  \bibnamefont {Nowak}}\ and\ \bibinfo {author} {\bibfnamefont {Katarzyna}\
  \bibnamefont {Sznajd-Weron}},\ }\bibfield  {title} {\enquote {\bibinfo
  {title} {Switching from a continuous to a discontinuous phase transition
  under quenched disorder},}\ }\href {\doibase 10.1103/PhysRevE.106.014125}
  {\bibfield  {journal} {\bibinfo  {journal} {Phys. Rev. E}\ }\textbf {\bibinfo
  {volume} {106}},\ \bibinfo {pages} {014125} (\bibinfo {year}
  {2022})}\BibitemShut {NoStop}%
\bibitem [{\citenamefont {J\k{e}drzejewski}\ and\ \citenamefont
  {Sznajd-Weron}(2017)}]{Jed:Szn:17}%
  \BibitemOpen
  \bibfield  {author} {\bibinfo {author} {\bibfnamefont {A.}~\bibnamefont
  {J\k{e}drzejewski}}\ and\ \bibinfo {author} {\bibfnamefont {K.}~\bibnamefont
  {Sznajd-Weron}},\ }\bibfield  {title} {\enquote {\bibinfo {title}
  {Person-situation debate revisited: Phase transitions with quenched and
  annealed disorders},}\ }\href {\doibase 10.3390/e19080415} {\bibfield
  {journal} {\bibinfo  {journal} {Entropy}\ }\textbf {\bibinfo {volume} {19}},\
  \bibinfo {pages} {415} (\bibinfo {year} {2017})}\BibitemShut {NoStop}%
\bibitem [{\citenamefont {Boccaletti}\ \emph {et~al.}(2006)\citenamefont
  {Boccaletti}, \citenamefont {Latora}, \citenamefont {Moreno}, \citenamefont
  {Chavez},\ and\ \citenamefont {Hwang}}]{Boc:etal:06}%
  \BibitemOpen
  \bibfield  {author} {\bibinfo {author} {\bibfnamefont {S.}~\bibnamefont
  {Boccaletti}}, \bibinfo {author} {\bibfnamefont {V.}~\bibnamefont {Latora}},
  \bibinfo {author} {\bibfnamefont {Y.}~\bibnamefont {Moreno}}, \bibinfo
  {author} {\bibfnamefont {M.}~\bibnamefont {Chavez}}, \ and\ \bibinfo {author}
  {\bibfnamefont {D.-U.}\ \bibnamefont {Hwang}},\ }\bibfield  {title} {\enquote
  {\bibinfo {title} {Complex networks: Structure and dynamics},}\ }\href
  {\doibase 10.1016/j.physrep.2005.10.009} {\bibfield  {journal} {\bibinfo
  {journal} {Phys. Rep.}\ }\textbf {\bibinfo {volume} {424}},\ \bibinfo {pages}
  {175--308} (\bibinfo {year} {2006})}\BibitemShut {NoStop}%
\bibitem [{\citenamefont {Nyczka}\ and\ \citenamefont
  {Sznajd-Weron}(2013)}]{Nyc:Szn:13}%
  \BibitemOpen
  \bibfield  {author} {\bibinfo {author} {\bibfnamefont {P.}~\bibnamefont
  {Nyczka}}\ and\ \bibinfo {author} {\bibfnamefont {K.}~\bibnamefont
  {Sznajd-Weron}},\ }\bibfield  {title} {\enquote {\bibinfo {title}
  {Anticonformity or independence?--{I}nsights from statistical physics},}\
  }\href {\doibase 10.1007/s10955-013-0701-4} {\bibfield  {journal} {\bibinfo
  {journal} {J. Stat. Phys.}\ }\textbf {\bibinfo {volume} {151}},\ \bibinfo
  {pages} {174--202} (\bibinfo {year} {2013})}\BibitemShut {NoStop}%
\bibitem [{\citenamefont {Nyczka}\ \emph {et~al.}(2012)\citenamefont {Nyczka},
  \citenamefont {Sznajd-Weron},\ and\ \citenamefont
  {Cis\l{}o}}]{Nyc:Szn:Cis:12}%
  \BibitemOpen
  \bibfield  {author} {\bibinfo {author} {\bibfnamefont {P.}~\bibnamefont
  {Nyczka}}, \bibinfo {author} {\bibfnamefont {K.}~\bibnamefont
  {Sznajd-Weron}}, \ and\ \bibinfo {author} {\bibfnamefont {J.}~\bibnamefont
  {Cis\l{}o}},\ }\bibfield  {title} {\enquote {\bibinfo {title} {Phase
  transitions in the $q$-voter model with two types of stochastic driving},}\
  }\href {\doibase 10.1103/PhysRevE.86.011105} {\bibfield  {journal} {\bibinfo
  {journal} {Phys. Rev. E}\ }\textbf {\bibinfo {volume} {86}},\ \bibinfo
  {pages} {011105} (\bibinfo {year} {2012})}\BibitemShut {NoStop}%
\bibitem [{\citenamefont {Castellano}\ \emph
  {et~al.}(2009{\natexlab{b}})\citenamefont {Castellano}, \citenamefont
  {Mu\~noz},\ and\ \citenamefont {Pastor-Satorras}}]{Cas:Mun:Pas:09}%
  \BibitemOpen
  \bibfield  {author} {\bibinfo {author} {\bibfnamefont {C.}~\bibnamefont
  {Castellano}}, \bibinfo {author} {\bibfnamefont {M.~A.}\ \bibnamefont
  {Mu\~noz}}, \ and\ \bibinfo {author} {\bibfnamefont {R.}~\bibnamefont
  {Pastor-Satorras}},\ }\bibfield  {title} {\enquote {\bibinfo {title}
  {Nonlinear $q$-voter model},}\ }\href {\doibase 10.1103/PhysRevE.80.041129}
  {\bibfield  {journal} {\bibinfo  {journal} {Phys. Rev. E}\ }\textbf {\bibinfo
  {volume} {80}},\ \bibinfo {pages} {041129} (\bibinfo {year}
  {2009}{\natexlab{b}})}\BibitemShut {NoStop}%
\bibitem [{\citenamefont {Grabisch}\ and\ \citenamefont
  {Rusinowska}(2020)}]{Gra:Rus:20}%
  \BibitemOpen
  \bibfield  {author} {\bibinfo {author} {\bibfnamefont {M.}~\bibnamefont
  {Grabisch}}\ and\ \bibinfo {author} {\bibfnamefont {A.}~\bibnamefont
  {Rusinowska}},\ }\bibfield  {title} {\enquote {\bibinfo {title} {A survey on
  nonstrategic models of opinion dynamics},}\ }\href {\doibase
  10.3390/g11040065} {\bibfield  {journal} {\bibinfo  {journal} {Games}\
  }\textbf {\bibinfo {volume} {11}},\ \bibinfo {pages} {65} (\bibinfo {year}
  {2020})}\BibitemShut {NoStop}%
\bibitem [{\citenamefont {Javarone}\ and\ \citenamefont
  {Squartini}(2015)}]{Jav:Squ:15}%
  \BibitemOpen
  \bibfield  {author} {\bibinfo {author} {\bibfnamefont {M.~A.}\ \bibnamefont
  {Javarone}}\ and\ \bibinfo {author} {\bibfnamefont {T.}~\bibnamefont
  {Squartini}},\ }\bibfield  {title} {\enquote {\bibinfo {title}
  {Conformism-driven phases of opinion formation on heterogeneous networks: the
  $q$-voter model case},}\ }\href {\doibase 10.1088/1742-5468/2015/10/P10002}
  {\bibfield  {journal} {\bibinfo  {journal} {J. Stat. Mech.: Theory Exp.}\
  }\textbf {\bibinfo {volume} {2015}},\ \bibinfo {pages} {P10002} (\bibinfo
  {year} {2015})}\BibitemShut {NoStop}%
\bibitem [{\citenamefont {Khalil}\ and\ \citenamefont
  {Toral}(2019)}]{Kha:Tor:19}%
  \BibitemOpen
  \bibfield  {author} {\bibinfo {author} {\bibfnamefont {N.}~\bibnamefont
  {Khalil}}\ and\ \bibinfo {author} {\bibfnamefont {R.}~\bibnamefont {Toral}},\
  }\bibfield  {title} {\enquote {\bibinfo {title} {The noisy voter model under
  the influence of contrarians},}\ }\href {\doibase
  10.1016/j.physa.2018.09.178} {\bibfield  {journal} {\bibinfo  {journal}
  {Physica A}\ }\textbf {\bibinfo {volume} {515}},\ \bibinfo {pages} {81--92}
  (\bibinfo {year} {2019})}\BibitemShut {NoStop}%
\bibitem [{\citenamefont {Javarone}(2014)}]{Jav:14}%
  \BibitemOpen
  \bibfield  {author} {\bibinfo {author} {\bibfnamefont {M.~A.}\ \bibnamefont
  {Javarone}},\ }\bibfield  {title} {\enquote {\bibinfo {title} {Social
  influences in opinion dynamics: The role of conformity},}\ }\href {\doibase
  10.1016/j.physa.2014.07.018} {\bibfield  {journal} {\bibinfo  {journal}
  {Physica A}\ }\textbf {\bibinfo {volume} {414}},\ \bibinfo {pages} {19--30}
  (\bibinfo {year} {2014})}\BibitemShut {NoStop}%
\bibitem [{\citenamefont {Vilela}\ \emph {et~al.}(2019)\citenamefont {Vilela},
  \citenamefont {Wang}, \citenamefont {Nelson},\ and\ \citenamefont
  {Stanley}}]{Vil:etal:19}%
  \BibitemOpen
  \bibfield  {author} {\bibinfo {author} {\bibfnamefont {A.~L.~M.}\
  \bibnamefont {Vilela}}, \bibinfo {author} {\bibfnamefont {C.}~\bibnamefont
  {Wang}}, \bibinfo {author} {\bibfnamefont {K.~P.}\ \bibnamefont {Nelson}}, \
  and\ \bibinfo {author} {\bibfnamefont {H.~E.}\ \bibnamefont {Stanley}},\
  }\bibfield  {title} {\enquote {\bibinfo {title} {Majority-vote model for
  financial markets},}\ }\href {\doibase 10.1016/j.physa.2018.10.007}
  {\bibfield  {journal} {\bibinfo  {journal} {Physica A}\ }\textbf {\bibinfo
  {volume} {515}},\ \bibinfo {pages} {762--770} (\bibinfo {year}
  {2019})}\BibitemShut {NoStop}%
\bibitem [{\citenamefont {Krawiecki}(2018)}]{Kra:18}%
  \BibitemOpen
  \bibfield  {author} {\bibinfo {author} {\bibfnamefont {A.}~\bibnamefont
  {Krawiecki}},\ }\bibfield  {title} {\enquote {\bibinfo {title}
  {Spin-glass-like transition in the majority-vote model with
  anticonformists},}\ }\href {\doibase 10.1140/epjb/e2018-80551-9} {\bibfield
  {journal} {\bibinfo  {journal} {Eur. Phys. J. B}\ }\textbf {\bibinfo {volume}
  {91}},\ \bibinfo {pages} {1--7} (\bibinfo {year} {2018})}\BibitemShut
  {NoStop}%
\bibitem [{\citenamefont {Stauffer}\ and\ \citenamefont {{Sá
  Martins}}(2004)}]{Sta:Sa:04}%
  \BibitemOpen
  \bibfield  {author} {\bibinfo {author} {\bibfnamefont {D.}~\bibnamefont
  {Stauffer}}\ and\ \bibinfo {author} {\bibfnamefont {J.S.}\ \bibnamefont {{Sá
  Martins}}},\ }\bibfield  {title} {\enquote {\bibinfo {title} {Simulation of
  {G}alam's contrarian opinions on percolative lattices},}\ }\href {\doibase
  https://doi.org/10.1016/j.physa.2003.12.003} {\bibfield  {journal} {\bibinfo
  {journal} {Physica A}\ }\textbf {\bibinfo {volume} {334}},\ \bibinfo {pages}
  {558--565} (\bibinfo {year} {2004})}\BibitemShut {NoStop}%
\bibitem [{\citenamefont {Baron}(2021{\natexlab{a}})}]{Bar:21a}%
  \BibitemOpen
  \bibfield  {author} {\bibinfo {author} {\bibfnamefont {J.~W.}\ \bibnamefont
  {Baron}},\ }\bibfield  {title} {\enquote {\bibinfo {title} {Persistent
  individual bias in a voter model with quenched disorder},}\ }\href {\doibase
  10.1103/PhysRevE.103.052309} {\bibfield  {journal} {\bibinfo  {journal}
  {Phys. Rev. E}\ }\textbf {\bibinfo {volume} {103}},\ \bibinfo {pages}
  {052309} (\bibinfo {year} {2021}{\natexlab{a}})}\BibitemShut {NoStop}%
\bibitem [{\citenamefont {Baron}(2021{\natexlab{b}})}]{Bar:21b}%
  \BibitemOpen
  \bibfield  {author} {\bibinfo {author} {\bibfnamefont {J.~W.}\ \bibnamefont
  {Baron}},\ }\bibfield  {title} {\enquote {\bibinfo {title} {Consensus,
  polarization, and coexistence in a continuous opinion dynamics model with
  quenched disorder},}\ }\href {\doibase 10.1103/PhysRevE.104.044309}
  {\bibfield  {journal} {\bibinfo  {journal} {Phys. Rev. E}\ }\textbf {\bibinfo
  {volume} {104}},\ \bibinfo {pages} {044309} (\bibinfo {year}
  {2021}{\natexlab{b}})}\BibitemShut {NoStop}%
\bibitem [{\citenamefont {Krawiecki}(2020)}]{Kra:20}%
  \BibitemOpen
  \bibfield  {author} {\bibinfo {author} {\bibfnamefont {A.}~\bibnamefont
  {Krawiecki}},\ }\bibfield  {title} {\enquote {\bibinfo {title} {Ferromagnetic
  and spin-glass-like transition in the majority vote model on complete and
  random graphs},}\ }\href {\doibase 10.1140/epjb/e2020-10288-9} {\bibfield
  {journal} {\bibinfo  {journal} {Eur. Phys. J. B}\ }\textbf {\bibinfo {volume}
  {93}},\ \bibinfo {pages} {1--14} (\bibinfo {year} {2020})}\BibitemShut
  {NoStop}%
\bibitem [{\citenamefont {Krawiecki}(2021)}]{Kra:21}%
  \BibitemOpen
  \bibfield  {author} {\bibinfo {author} {\bibfnamefont {A.}~\bibnamefont
  {Krawiecki}},\ }\bibfield  {title} {\enquote {\bibinfo {title} {Ferromagnetic
  and spin-glass like transition in the $q$-neighbor ising model on random
  graphs},}\ }\href {\doibase 10.1140/epjb/s10051-021-00084-0} {\bibfield
  {journal} {\bibinfo  {journal} {Eur. Phys. J. B}\ }\textbf {\bibinfo {volume}
  {94}},\ \bibinfo {pages} {1--15} (\bibinfo {year} {2021})}\BibitemShut
  {NoStop}%
\bibitem [{\citenamefont {J\k{e}drzejewski}(2017)}]{Jed:17}%
  \BibitemOpen
  \bibfield  {author} {\bibinfo {author} {\bibfnamefont {A.}~\bibnamefont
  {J\k{e}drzejewski}},\ }\bibfield  {title} {\enquote {\bibinfo {title} {Pair
  approximation for the $q$-voter model with independence on complex
  networks},}\ }\href {\doibase 10.1103/PhysRevE.95.012307} {\bibfield
  {journal} {\bibinfo  {journal} {Phys. Rev. E}\ }\textbf {\bibinfo {volume}
  {95}},\ \bibinfo {pages} {012307} (\bibinfo {year} {2017})}\BibitemShut
  {NoStop}%
\bibitem [{\citenamefont {Peralta}\ \emph {et~al.}(2018)\citenamefont
  {Peralta}, \citenamefont {Carro}, \citenamefont {San~Miguel},\ and\
  \citenamefont {Toral}}]{Per:etal:18}%
  \BibitemOpen
  \bibfield  {author} {\bibinfo {author} {\bibfnamefont {A.~F.}\ \bibnamefont
  {Peralta}}, \bibinfo {author} {\bibfnamefont {A.}~\bibnamefont {Carro}},
  \bibinfo {author} {\bibfnamefont {M.}~\bibnamefont {San~Miguel}}, \ and\
  \bibinfo {author} {\bibfnamefont {R.}~\bibnamefont {Toral}},\ }\bibfield
  {title} {\enquote {\bibinfo {title} {Analytical and numerical study of the
  non-linear noisy voter model on complex networks},}\ }\href {\doibase
  10.1063/1.5030112} {\bibfield  {journal} {\bibinfo  {journal} {Chaos}\
  }\textbf {\bibinfo {volume} {28}},\ \bibinfo {pages} {075516} (\bibinfo
  {year} {2018})}\BibitemShut {NoStop}%
\bibitem [{\citenamefont {Gradowski}\ and\ \citenamefont
  {Krawiecki}(2020)}]{Gra:Kra:20}%
  \BibitemOpen
  \bibfield  {author} {\bibinfo {author} {\bibfnamefont {T.}~\bibnamefont
  {Gradowski}}\ and\ \bibinfo {author} {\bibfnamefont {A.}~\bibnamefont
  {Krawiecki}},\ }\bibfield  {title} {\enquote {\bibinfo {title} {Pair
  approximation for the $q$-voter model with independence on multiplex
  networks},}\ }\href {\doibase 10.1103/PhysRevE.102.022314} {\bibfield
  {journal} {\bibinfo  {journal} {Phys. Rev. E}\ }\textbf {\bibinfo {volume}
  {102}},\ \bibinfo {pages} {022314} (\bibinfo {year} {2020})}\BibitemShut
  {NoStop}%
\bibitem [{\citenamefont {J\k{e}drzejewski}\ \emph {et~al.}(2020)\citenamefont
  {J\k{e}drzejewski}, \citenamefont {Nowak}, \citenamefont {Abramiuk},\ and\
  \citenamefont {Sznajd-Weron}}]{Jed:etal:20}%
  \BibitemOpen
  \bibfield  {author} {\bibinfo {author} {\bibfnamefont {A.}~\bibnamefont
  {J\k{e}drzejewski}}, \bibinfo {author} {\bibfnamefont {B.}~\bibnamefont
  {Nowak}}, \bibinfo {author} {\bibfnamefont {A.}~\bibnamefont {Abramiuk}}, \
  and\ \bibinfo {author} {\bibfnamefont {K.}~\bibnamefont {Sznajd-Weron}},\
  }\bibfield  {title} {\enquote {\bibinfo {title} {Competing local and global
  interactions in social dynamics: How important is the friendship network?}}\
  }\href {\doibase 10.1063/5.0004797} {\bibfield  {journal} {\bibinfo
  {journal} {Chaos}\ }\textbf {\bibinfo {volume} {30}},\ \bibinfo {pages}
  {073105} (\bibinfo {year} {2020})}\BibitemShut {NoStop}%
\bibitem [{\citenamefont {Abramiuk}\ and\ \citenamefont
  {Sznajd-Weron}(2020)}]{Abr:Szn:20}%
  \BibitemOpen
  \bibfield  {author} {\bibinfo {author} {\bibfnamefont {A.}~\bibnamefont
  {Abramiuk}}\ and\ \bibinfo {author} {\bibfnamefont {K.}~\bibnamefont
  {Sznajd-Weron}},\ }\bibfield  {title} {\enquote {\bibinfo {title}
  {Generalized independence in the $q$-voter model: How do parameters influence
  the phase transition?}}\ }\href {\doibase 10.3390/e22010120} {\bibfield
  {journal} {\bibinfo  {journal} {Entropy}\ }\textbf {\bibinfo {volume} {22}},\
  \bibinfo {pages} {120} (\bibinfo {year} {2020})}\BibitemShut {NoStop}%
\bibitem [{\citenamefont {Vieira}\ \emph {et~al.}(2020)\citenamefont {Vieira},
  \citenamefont {Peralta}, \citenamefont {Toral}, \citenamefont {Miguel},\ and\
  \citenamefont {Anteneodo}}]{Vie:etal:20}%
  \BibitemOpen
  \bibfield  {author} {\bibinfo {author} {\bibfnamefont {A.~R.}\ \bibnamefont
  {Vieira}}, \bibinfo {author} {\bibfnamefont {A.~F.}\ \bibnamefont {Peralta}},
  \bibinfo {author} {\bibfnamefont {R.}~\bibnamefont {Toral}}, \bibinfo
  {author} {\bibfnamefont {M.~San}\ \bibnamefont {Miguel}}, \ and\ \bibinfo
  {author} {\bibfnamefont {C.}~\bibnamefont {Anteneodo}},\ }\bibfield  {title}
  {\enquote {\bibinfo {title} {Pair approximation for the noisy threshold
  $q$-voter model},}\ }\href {\doibase 10.1103/PhysRevE.101.052131} {\bibfield
  {journal} {\bibinfo  {journal} {Phys. Rev. E}\ }\textbf {\bibinfo {volume}
  {101}},\ \bibinfo {pages} {052131} (\bibinfo {year} {2020})}\BibitemShut
  {NoStop}%
\bibitem [{\citenamefont {Abramiuk-Szurlej}\ \emph {et~al.}(2021)\citenamefont
  {Abramiuk-Szurlej}, \citenamefont {Lipiecki}, \citenamefont {Paw{\l}owski},\
  and\ \citenamefont {Sznajd-Weron}}]{Abr:etal:21}%
  \BibitemOpen
  \bibfield  {author} {\bibinfo {author} {\bibfnamefont {A.}~\bibnamefont
  {Abramiuk-Szurlej}}, \bibinfo {author} {\bibfnamefont {A.}~\bibnamefont
  {Lipiecki}}, \bibinfo {author} {\bibfnamefont {J.}~\bibnamefont
  {Paw{\l}owski}}, \ and\ \bibinfo {author} {\bibfnamefont {K.}~\bibnamefont
  {Sznajd-Weron}},\ }\bibfield  {title} {\enquote {\bibinfo {title}
  {Discontinuous phase transitions in the $q$-voter model with generalized
  anticonformity on random graphs},}\ }\href {\doibase
  10.1038/s41598-021-97155-0} {\bibfield  {journal} {\bibinfo  {journal} {Sci.
  Rep.}\ }\textbf {\bibinfo {volume} {11}},\ \bibinfo {pages} {1--9} (\bibinfo
  {year} {2021})}\BibitemShut {NoStop}%
\bibitem [{\citenamefont {Schweitzer}\ and\ \citenamefont
  {Behera}(2009)}]{Sch:Beh:09}%
  \BibitemOpen
  \bibfield  {author} {\bibinfo {author} {\bibfnamefont {F.}~\bibnamefont
  {Schweitzer}}\ and\ \bibinfo {author} {\bibfnamefont {L.}~\bibnamefont
  {Behera}},\ }\bibfield  {title} {\enquote {\bibinfo {title} {Nonlinear voter
  models: the transition from invasion to coexistence},}\ }\href {\doibase
  10.1140/epjb/e2009-00001-3} {\bibfield  {journal} {\bibinfo  {journal} {Eur.
  Phys. J. B}\ }\textbf {\bibinfo {volume} {67}},\ \bibinfo {pages} {301--318}
  (\bibinfo {year} {2009})}\BibitemShut {NoStop}%
\bibitem [{\citenamefont {Gleeson}(2013)}]{Gle:13}%
  \BibitemOpen
  \bibfield  {author} {\bibinfo {author} {\bibfnamefont {James~P.}\
  \bibnamefont {Gleeson}},\ }\bibfield  {title} {\enquote {\bibinfo {title}
  {Binary-state dynamics on complex networks: Pair approximation and beyond},}\
  }\href {\doibase 10.1103/PhysRevX.3.021004} {\bibfield  {journal} {\bibinfo
  {journal} {Phys. Rev. X}\ }\textbf {\bibinfo {volume} {3}},\ \bibinfo {pages}
  {021004} (\bibinfo {year} {2013})}\BibitemShut {NoStop}%
\bibitem [{\citenamefont {Chmiel}\ \emph {et~al.}(2018)\citenamefont {Chmiel},
  \citenamefont {Gradowski},\ and\ \citenamefont {Krawiecki}}]{Chm:Gra:Kra:18}%
  \BibitemOpen
  \bibfield  {author} {\bibinfo {author} {\bibfnamefont {A.}~\bibnamefont
  {Chmiel}}, \bibinfo {author} {\bibfnamefont {T.}~\bibnamefont {Gradowski}}, \
  and\ \bibinfo {author} {\bibfnamefont {A.}~\bibnamefont {Krawiecki}},\
  }\bibfield  {title} {\enquote {\bibinfo {title} {$q$-neighbor ising model on
  random networks},}\ }\href {\doibase 10.1142/S0129183118500419} {\bibfield
  {journal} {\bibinfo  {journal} {Int. J. Mod. Phys. A}\ }\textbf {\bibinfo
  {volume} {29}},\ \bibinfo {pages} {1850041} (\bibinfo {year}
  {2018})}\BibitemShut {NoStop}%
\bibitem [{\citenamefont {Peralta}\ and\ \citenamefont
  {Toral}(2020)}]{Per:Tor:20}%
  \BibitemOpen
  \bibfield  {author} {\bibinfo {author} {\bibfnamefont {A.~F.}\ \bibnamefont
  {Peralta}}\ and\ \bibinfo {author} {\bibfnamefont {R.}~\bibnamefont
  {Toral}},\ }\bibfield  {title} {\enquote {\bibinfo {title} {Binary-state
  dynamics on complex networks: Stochastic pair approximation and beyond},}\
  }\href {\doibase 10.1103/PhysRevResearch.2.043370} {\bibfield  {journal}
  {\bibinfo  {journal} {Phys. Rev. Research}\ }\textbf {\bibinfo {volume}
  {2}},\ \bibinfo {pages} {043370} (\bibinfo {year} {2020})}\BibitemShut
  {NoStop}%
\bibitem [{\citenamefont {Diakonova}\ \emph {et~al.}(2014)\citenamefont
  {Diakonova}, \citenamefont {San~Miguel},\ and\ \citenamefont
  {Egu\'{\i}luz}}]{Dia:San:Egu:14}%
  \BibitemOpen
  \bibfield  {author} {\bibinfo {author} {\bibfnamefont {M.}~\bibnamefont
  {Diakonova}}, \bibinfo {author} {\bibfnamefont {M.}~\bibnamefont
  {San~Miguel}}, \ and\ \bibinfo {author} {\bibfnamefont {V.~M.}\ \bibnamefont
  {Egu\'{\i}luz}},\ }\bibfield  {title} {\enquote {\bibinfo {title} {Absorbing
  and shattered fragmentation transitions in multilayer coevolution},}\ }\href
  {\doibase 10.1103/PhysRevE.89.062818} {\bibfield  {journal} {\bibinfo
  {journal} {Phys. Rev. E}\ }\textbf {\bibinfo {volume} {89}},\ \bibinfo
  {pages} {062818} (\bibinfo {year} {2014})}\BibitemShut {NoStop}%
\bibitem [{\citenamefont {Min}\ and\ \citenamefont
  {San~Miguel}(2017)}]{Min:San:17}%
  \BibitemOpen
  \bibfield  {author} {\bibinfo {author} {\bibfnamefont {B.}~\bibnamefont
  {Min}}\ and\ \bibinfo {author} {\bibfnamefont {M.}~\bibnamefont
  {San~Miguel}},\ }\bibfield  {title} {\enquote {\bibinfo {title}
  {Fragmentation transitions in a coevolving nonlinear voter model},}\ }\href
  {\doibase 10.1038/s41598-017-13047-2} {\bibfield  {journal} {\bibinfo
  {journal} {Sci. Rep.}\ }\textbf {\bibinfo {volume} {7}},\ \bibinfo {pages}
  {1--9} (\bibinfo {year} {2017})}\BibitemShut {NoStop}%
\bibitem [{\citenamefont {Toruniewska}\ \emph {et~al.}(2017)\citenamefont
  {Toruniewska}, \citenamefont {Ku{\l}akowski}, \citenamefont {Suchecki},\ and\
  \citenamefont {Ho{\l}yst}}]{Tor:etal:17}%
  \BibitemOpen
  \bibfield  {author} {\bibinfo {author} {\bibfnamefont {J.}~\bibnamefont
  {Toruniewska}}, \bibinfo {author} {\bibfnamefont {K.}~\bibnamefont
  {Ku{\l}akowski}}, \bibinfo {author} {\bibfnamefont {K.}~\bibnamefont
  {Suchecki}}, \ and\ \bibinfo {author} {\bibfnamefont {J.~A.}\ \bibnamefont
  {Ho{\l}yst}},\ }\bibfield  {title} {\enquote {\bibinfo {title} {Coupling of
  link-and node-ordering in the coevolving voter model},}\ }\href {\doibase
  10.1103/PhysRevE.96.042306} {\bibfield  {journal} {\bibinfo  {journal} {Phys.
  Rev. E}\ }\textbf {\bibinfo {volume} {96}},\ \bibinfo {pages} {042306}
  (\bibinfo {year} {2017})}\BibitemShut {NoStop}%
\bibitem [{\citenamefont {J\ifmmode~\mbox{\k{e}}\else \k{e}\fi{}drzejewski}\
  \emph {et~al.}(2020)\citenamefont {J\ifmmode~\mbox{\k{e}}\else
  \k{e}\fi{}drzejewski}, \citenamefont {Toruniewska}, \citenamefont {Suchecki},
  \citenamefont {Zaikin},\ and\ \citenamefont {Ho\l{}yst}}]{Jed:Tor:etal:20}%
  \BibitemOpen
  \bibfield  {author} {\bibinfo {author} {\bibfnamefont {A.}~\bibnamefont
  {J\ifmmode~\mbox{\k{e}}\else \k{e}\fi{}drzejewski}}, \bibinfo {author}
  {\bibfnamefont {J.}~\bibnamefont {Toruniewska}}, \bibinfo {author}
  {\bibfnamefont {K.}~\bibnamefont {Suchecki}}, \bibinfo {author}
  {\bibfnamefont {O.}~\bibnamefont {Zaikin}}, \ and\ \bibinfo {author}
  {\bibfnamefont {J.~A.}\ \bibnamefont {Ho\l{}yst}},\ }\bibfield  {title}
  {\enquote {\bibinfo {title} {Spontaneous symmetry breaking of active phase in
  coevolving nonlinear voter model},}\ }\href {\doibase
  10.1103/PhysRevE.102.042313} {\bibfield  {journal} {\bibinfo  {journal}
  {Phys. Rev. E}\ }\textbf {\bibinfo {volume} {102}},\ \bibinfo {pages}
  {042313} (\bibinfo {year} {2020})}\BibitemShut {NoStop}%
\bibitem [{\citenamefont {Raducha}\ and\ \citenamefont
  {San~Miguel}(2020)}]{Rad:San:20}%
  \BibitemOpen
  \bibfield  {author} {\bibinfo {author} {\bibfnamefont {T.}~\bibnamefont
  {Raducha}}\ and\ \bibinfo {author} {\bibfnamefont {M.}~\bibnamefont
  {San~Miguel}},\ }\bibfield  {title} {\enquote {\bibinfo {title} {Emergence of
  complex structures from nonlinear interactions and noise in coevolving
  networks},}\ }\href {\doibase 10.1038/s41598-020-72662-8} {\bibfield
  {journal} {\bibinfo  {journal} {Sci. Rep.}\ }\textbf {\bibinfo {volume}
  {10}},\ \bibinfo {pages} {1--14} (\bibinfo {year} {2020})}\BibitemShut
  {NoStop}%
\bibitem [{\citenamefont {Vazquez}\ and\ \citenamefont
  {Egu{\'\i}luz}(2008)}]{Vaz:Egu:08}%
  \BibitemOpen
  \bibfield  {author} {\bibinfo {author} {\bibfnamefont {F.}~\bibnamefont
  {Vazquez}}\ and\ \bibinfo {author} {\bibfnamefont {V.~M.}\ \bibnamefont
  {Egu{\'\i}luz}},\ }\bibfield  {title} {\enquote {\bibinfo {title} {Analytical
  solution of the voter model on uncorrelated networks},}\ }\href {\doibase
  10.1088/1367-2630/10/6/063011} {\bibfield  {journal} {\bibinfo  {journal}
  {New J. Phys.}\ }\textbf {\bibinfo {volume} {10}},\ \bibinfo {pages} {063011}
  (\bibinfo {year} {2008})}\BibitemShut {NoStop}%
\bibitem [{\citenamefont {Galam}(2004)}]{Gal:04}%
  \BibitemOpen
  \bibfield  {author} {\bibinfo {author} {\bibfnamefont {S.}~\bibnamefont
  {Galam}},\ }\bibfield  {title} {\enquote {\bibinfo {title} {Contrarian
  deterministic effects on opinion dynamics: “the hung elections
  scenario”},}\ }\href {\doibase https://doi.org/10.1016/j.physa.2003.10.041}
  {\bibfield  {journal} {\bibinfo  {journal} {Physica A}\ }\textbf {\bibinfo
  {volume} {333}},\ \bibinfo {pages} {453--460} (\bibinfo {year}
  {2004})}\BibitemShut {NoStop}%
\bibitem [{\citenamefont {Ruci\'{n}ski}\ and\ \citenamefont
  {Wormald}(1992)}]{Ruc:Wor:92}%
  \BibitemOpen
  \bibfield  {author} {\bibinfo {author} {\bibfnamefont {A.}~\bibnamefont
  {Ruci\'{n}ski}}\ and\ \bibinfo {author} {\bibfnamefont {N.~C.}\ \bibnamefont
  {Wormald}},\ }\bibfield  {title} {\enquote {\bibinfo {title} {Random graph
  processes with degree restrictions},}\ }\href {\doibase
  10.1017/S0963548300000183} {\bibfield  {journal} {\bibinfo  {journal} {Comb.
  Probab. Comput.}\ }\textbf {\bibinfo {volume} {1}},\ \bibinfo {pages}
  {169–180} (\bibinfo {year} {1992})}\BibitemShut {NoStop}%
\bibitem [{\citenamefont {Barab{\'a}si}\ and\ \citenamefont
  {Albert}(1999)}]{Bar:Alb:99}%
  \BibitemOpen
  \bibfield  {author} {\bibinfo {author} {\bibfnamefont {A.-L.}\ \bibnamefont
  {Barab{\'a}si}}\ and\ \bibinfo {author} {\bibfnamefont {R.}~\bibnamefont
  {Albert}},\ }\bibfield  {title} {\enquote {\bibinfo {title} {Emergence of
  scaling in random networks},}\ }\href {\doibase 10.1126/science.286.5439.509}
  {\bibfield  {journal} {\bibinfo  {journal} {Science}\ }\textbf {\bibinfo
  {volume} {286}},\ \bibinfo {pages} {509--512} (\bibinfo {year}
  {1999})}\BibitemShut {NoStop}%
\end{thebibliography}%
\end{document}